\documentclass[fleqn,usenatbib]{mnras}

\usepackage{newtxtext,newtxmath}
\usepackage[T1]{fontenc}
\DeclareRobustCommand{\VAN}[3]{#2}
\let\VANthebibliography\thebibliography
\def\thebibliography{\DeclareRobustCommand{\VAN}[3]{##3}\VANthebibliography}

\usepackage[utf8x]{inputenc}
\usepackage{graphicx}	% Including figure files
\usepackage{amsmath}	% Advanced maths commands
\usepackage{esint}
\usepackage{academicons}
\usepackage{xcolor}
\usepackage{orcidlink}
\usepackage{siunitx}
\usepackage{blindtext}
\usepackage{multicol}
\usepackage{multirow}
\usepackage{subfig}
\usepackage{hyperref}
\usepackage{orcidlink}
\usepackage{natbib}
\usepackage{comment}
\usepackage{bm}
\usepackage{lmodern}
\usepackage{anyfontsize} 
\usepackage{tensor}
\usepackage{leftindex}
\DeclarePairedDelimiterXPP\BigOSI[2]%
  {\mathcal{O}}{(}{)}{}%
  {\SI{#1}{#2}}

\title[Optimal ICs of $\nu$HDM]{On the initial conditions of the $\nu$HDM cosmological model} 

\author[N. Samaras et al.]
{Nick Samaras$^{1}$ \thanks{E-mail: nicksam@sirrah.troja.mff.cuni.cz} $^{\orcidlink{0000-0001-8375-6652}}$,
Sebastian Grandis $^{3}$ $^{\orcidlink{0000-0002-4577-8217}}$
and Pavel Kroupa $^{1,2}$ $^{\orcidlink{0000-0002-7301-3377}}$
\\
$^{1}$ Astronomical Institute, Faculty of Mathematics and Physics, Charles University, V Hole\v{s}ovi\v{c}k\'ach 2, CZ-180 00 Praha 8, Czech Republic \\
$^{2}$ Helmholtz-Institut f\"ur Strahlen und Kernphysik (HISKP), University of Bonn, Nussallee 14−16, D 53115 Bonn,Germany\\
$^{3}$ Universit\"at Innsbruck, Institut f\"ur Astro- und Teilchenphysik, Technikerstr. 25/8, A-6020 Innsbruck, Austria}

\date{Accepted XXX. Received YYY; in original form ZZZ}
\pubyear{2024}
\DeclareUnicodeCharacter{2212}{-}
\begin{document}
\label{firstpage}
\pagerange{\pageref{firstpage}--\pageref{lastpage}}
\maketitle

% Abstract of the paper
\begin{abstract}
The $\nu$HDM is the only cosmological model based on Milgromian Dynamics (MOND) with available structure formation simulations. While MOND accounts for galaxies, with a priori predictions for spirals and ellipticals, a light sterile neutrino of 11 eV can assist in recovering scaling relations on the galaxy-cluster scales.
%and the CMB power spectrum. The structure formation simulations, published in \citet{Witt}, although based on the Planck parameters of $\omega_{b} h^2 = 0.2238$, $\Omega_{m} = 0.31 $ and $H_0 = 67.4$ km/sec/Mpc, lacked a cosmological likelihood-based inference. 
In order to perform MONDian cosmological simulations in this theoretical approach, initial conditions derived from a fit to the angular power spectrum of Cosmic Microwave Background (CMB) fluctuations are required. In this work, we employ CosmoSIS to perform a Bayesian study of the $\nu$HDM model. Using the best-fit values of the posterior, the CMB power spectrum is reevaluated. The excess of power in the transfer function implies a distinct evolution scenario, which can be used further as an input for a set of hydro-dynamical calculations. The resulting values $H_{0} \approx 56$ km/s/Mpc and $\Omega_{m_0}\approx$ 0.5 are far from agreement with respect to the best fit ones in the canonical Cold Dark Matter model, but may be significant in MONDian cosmology. The assumed \textit{Planck} CMB initial conditions are only valid for the $\Lambda$CDM cosmology. This work constitutes a first step in an iterative procedure needed to disentangle the model dependence of the derived initial density and velocity fields.
\vspace{0.5cm}
\end{abstract}

% Select between one and six entries from the list of approved keywords.
% Don't make up new ones.
\begin{keywords}
Cosmology -- gravitation -- galaxies: statistics
\end{keywords}

%%%%%%%%%%%%%%%%%%%%%%%%%%%%%%%%%%%%%%%%%%%%%%%%%%

%%%%%%%%%%%%%%%%% BODY OF PAPER %%%%%%%%%%%%%%%%%%

\section{Introduction}
\begin{comment}
The current standard representative picture of the birth and evolution of the universe is summarised in the so-called $\Lambda$CDM scenario. That is, the model of a Big-Bang (BB) beginning of the cosmos, approximately 13.7 Gyr ago and the main players shaping its progress being unknown exotic components like Dark Matter and Dark Energy. While the former compensates for the galactic rotation to what Newtonian Gravity is missing, the latter accounts for the late-time apparent acceleration of the expansion of the universe. Although this synopsis offers plenty of explanations on the behaviour of the galactic systems, it suffers from a number of open questions. From the recently established "Hubble tension", the disagreement of different measurements regarding the pace at which the Universe is stretching outwards, to the an-isotropic distribution of smaller satellite galaxies around the bigger -parent- Milky way and Andromeda galaxy, the theory seems incomplete from many aspects.
\end{comment}

%SG: hi all, I would propose rephrasing the introduction in a more generic way, highlighting the observational data and analysis techniques of modern cosmology. In my opinion, these are independent on the specific cosmological model and provide a good introduction and motivation to the work Niko undertook.

Over the last decades, cosmology has moved from a theoretical discipline to a data-driven one. This transition has come about in no small part due to the unprecedented precision of extra-galactic observation, chief among them the measurement of the temperature anisotropies in the Cosmic Microwave Background (CMB) carried out by the ESA mission \textit{Planck} \citep[see][for an overview]{2020A&A...641A...1P}. Further significant contributions to observational cosmology came from galaxy redshift surveys \citep{2025JCAP...02..021A} and supernovae Type Ia observations (SNe Ia) \citep{2022ApJ...934L...7R}. These observations have recently been complemented by weak lensing \citep{2024A&A...683A.240J} and photometric galaxy surveys, CMB lensing measurements, and cluster surveys that probe the Large Scale Structure (LSS) of the Universe, namely, its matter distribution at late times.

These large data volumes are compressed into so-called summary statistics, which are then fitted by theory predictions using Bayesian inference frameworks. The CMB anisotropies are compressed into their angular power spectrum, which can be accounted for by the underlying cosmological model using our laboratory-tested knowledge of plasma physics. Galaxy redshift surveys (see for instance the SDSS \citep{2000AJ....120.1579Y}) are compressed into measurements of the angular scale of Baryonic Acoustic Oscillations (BAO), which provide empirical constraints on the distance-redshift relation, as do SNe Ia. This relation can be obtained by assuming a metric theory of gravity and by considering the Universe to be isotropic and homogeneous on large scales. Analytical calculations for LSS observables only reach an accuracy of $\approx 10\%$ and are therefore supplemented with large-volume cosmological simulations to reach the required percent-level accuracy. These simulations require initial conditions (ICs), which are taken from the best-fit results of the CMB and the distance measures. The primary observational benchmark for any quantitative theory of cosmology is therefore its ability to accurately fit the angular power spectrum of the CMB.

On a purely phenomenological level, the peaks in the angular power spectrum of the CMB show the presence of acoustic waves in the primordial plasma. Such behavior is naturally expected through the tight interactions between the electron and the photons in the plasma. Intriguingly, the peak height of the even-numbered peaks is smaller than that of the odd-numbered peaks once one accounts for the damping of higher-frequency oscillations. This has been interpreted as the presence of a gravitational potential sourced by a component that does not interact with the photons, which can thus be agnostically described as 'dark' (in the sense that it does not interact with photons). The standard model of cosmology proposes a hitherto undetected Cold Dark Matter (CDM) for this component, as opposed to a hot or warm component. While this enables accurate prediction for the LSS observations, it has run into several issues on small scales \citep{2023eppg.confE.231K}. As no suitable particle for CDM exists in the standard model of particle physics, significant reservations about this assumption persist. Modified Newtonian Dynamics, or Milgromian Dynamics (hereafter MOND) is an alternative to the CDM model, which does not assume a cold dark matter component but enhances gravity for weak gravitational accelerations.
%SGAmongst the numerous alternatives of $\Lambda$CDM \citep{1990Natur.348..705E, 1995Natur.377..600O}

In the absence of CDM, another dark component is required to reproduce the correct phenomenology of the CMB power spectrum. One proposed scenario is the $\nu$HDM cosmological model \citep{10.1111/j.1365-2966.2011.19321.x, Katz_2013, 10.1093/mnras/staa2348, Witt}, which relies on MOND and a hypothetical but yet-undetected sterile neutrino of $m_{\nu_{s}}$ = 11 eV/c\textsuperscript{2} $= 1.96 \times 10^{-32}$ g rest-mass. This rather conservative idea is still based on a Big Bang (BB) beginning of the Universe, assumes an inflationary era and adopts Dark Energy to explain the accelerated late-time expansion. Keeping the same evolution history of a Friedmann-Lemaitre-Robertson-Walker (FLRW) metric, the CDM part is substituted by an equal-contributing component to the total mass-energy budget by Hot Dark Matter (HDM). That is, at the current epoch for a $\Lambda$CDM Universe:
\begin{equation} 
    \begin{split} 
    \Omega_{m_{0}} & = \Omega_{\text{CDM}} \text{ + } \Omega_{\text{b}} \text{ + } \Omega_{\nu} \xrightarrow[\Lambda\text{CDM}]{\Omega_{\nu}\ll \Omega_\text{{CDM}}}\Omega_{\text{CDM}} \text{ + } \Omega_{\text{b}} \approx 0.26 \text{ + } 0.04,\\
    % \approx 0.315 & \xrightarrow[h = 0.6732]{\text{PLANCK}}\Omega_{\text{CDM}} h^2 +\Omega_{b} h^2 = 0.12011 + 0.022383 =  0.1431 \\
    \end{split}
\end{equation}
on the other hand, for an $\nu$HDM Universe:
\begin{equation} 
    \begin{split} 
    \Omega_{m_{0}} & = \Omega_{\text{CDM}}\text{ + }\Omega_{\text{b}}\text{ + }\Omega_{\nu} \xrightarrow[\nu\text{HDM}]{\Omega_{\nu}\gg\Omega_\text{{CDM}}}\Omega_{\nu}\text{ + }\Omega_{\text{b}} \approx 0.26\text{ + }0.04,\\
    %\approx 0.315 & \xrightarrow[]{\times h^2}=  0.1431 = \Omega_{\nu}h^2 +\Omega_{b} h^2 = 0.12011 + 0.022383 
    \end{split}
\end{equation}
where $\Omega_{m_{0}}$ is the total mass-energy density today ($\Omega_m$ for the rest of the paper), $\Omega_{\text{CDM}}$ is the Cold Dark Matter contribution, $\Omega_{\text{b}}$ is the normal baryonic matter and $\Omega_{\nu}$ is the neutrino contribution.

Any HDM approach washes out the structures on intermediate and small scales and is therefore unlikely to make even qualitatively accurate predictions of the LSS. In this context, the $\nu$HDM model is linked with MOND, which predicts stronger gravitational forces for very weak gravitational accelerations, potentially compensating for the effect of HDM. In the rest of the work, we will therefore assume that $\nu$HDM follows MOND for its small-scale gravitational physics. The viability of this model needs to be tested with dedicated cosmological simulations.

The latest study of the $\nu$HDM model, based on hydro-dynamical simulations by \citet{Witt}, has confirmed that structure formation behaves rather differently compared to the standard $\Lambda$CDM model, but most importantly that it is in tension with several of the new observations by the James Webb Space Telescope \citep{2023NatAs...7..622C, 2024arXiv240310238C}, that show galaxies to be forming at $10<z<20$. Gravitational bound objects appear quite late, at z $\approx$ 4, and the most massive clusters have a mean mass density (their total mass of baryons and neutrinos $M_{180} \equiv 180 {\rho}_{\text{crit}} \times \frac{4}{3}\pi R^{3}_{180}$, while $R_{180}$ is the radius enclosing that mass density) of $M_{180} \gtrsim 10^{17} M_{\odot}$ at $z=0$ for a 200 Mpc/h box \citep[their fig. 6]{Witt}. Furthermore, the mass function of the simulated galaxies is too shallow, leading to an overabundance of massive structures at the high mass end compared to the weak lensing catalogue of \citet{Wang_2022}. However, one should keep in mind that early galaxy formation, the observed existence of extremely massive galaxy clusters as well as Gpc-scale under-densities pose a serious problem for the broadly accepted $\Lambda$CDM model, too \citep{2021MNRAS.500.5249A, 2020MNRAS.499.2845H, Haslbauer_2022}.

In \citet{Witt}, the initial conditions of their simulations were set by varying the parameters of the $\nu$HDM model by hand to match the CMB power spectrum. In this work, we improve this procedure by utilizing the cosmological parameter estimation code, CosmoSIS \citep{2015A&C....12...45Z} \footnote{\url{https://cosmosis.readthedocs.io/en/latest/index.html}} to sample and determine the values that best fit the \textit{Planck} CMB power spectrum in $\nu$HDM cosmology. We underline that fact that the temperature fluctuations have to be corrected accordingly to the new model, which follows the MOND gravity and structure formation proceeds differently. An iterative procedure is imperative to accurately represent the changes with respect to the $\Lambda$CDM-estimated foregrounds. 

%published \textit{Planck} CMB power spectrum, we are performing a first step in what will have to be an iterative procedure, because the true ICs must take into account the different corrections for structure formation in opt-$\nu$HDM model compared to the standard $\Lambda$CDM model (Section \ref{corrections}).} 

%We have utilized the cosmological parameter estimation code CosmoSIS \citep{2015A&C....12...45Z} \footnote{\url{https://cosmosis.readthedocs.io/en/latest/index.html}} to re-sample and determine the values that govern the $\nu$HDM cosmology. After comparing the Likelihoods of \citet{Witt} paper to the $\Lambda$CDM concordance model, we performed a Markov Chain Monte Carlo (MCMC) \citep{10.2140/camcos.2010.5.65} simulation to match the CMB $C_{\ell}$ data. Despite of the improved CMB fit, the resulting cosmological parameters differ by construction from the $\Lambda$CDM inferred ones and radically from the latest DESI results too \citep{2024arXiv240403002D}.

The paper is organized as follows: in Section \ref{cosmo_model} we present the $\nu$HDM model; in Section \ref{cosmosis} CosmoSIS procedures are documented; in Section \ref{results} we summarize our results, comparing them to $\Lambda$CDM model; in Section \ref{Discussion} we elaborate the consequences of the resulting parameters and in Section \ref{conclusions}, we conclude our investigation.

\section{\texorpdfstring{The $\nu$HDM cosmological model}{The nuHDM cosmological model}}\label{cosmo_model}

\begin{figure} 
  \includegraphics[width=\columnwidth]{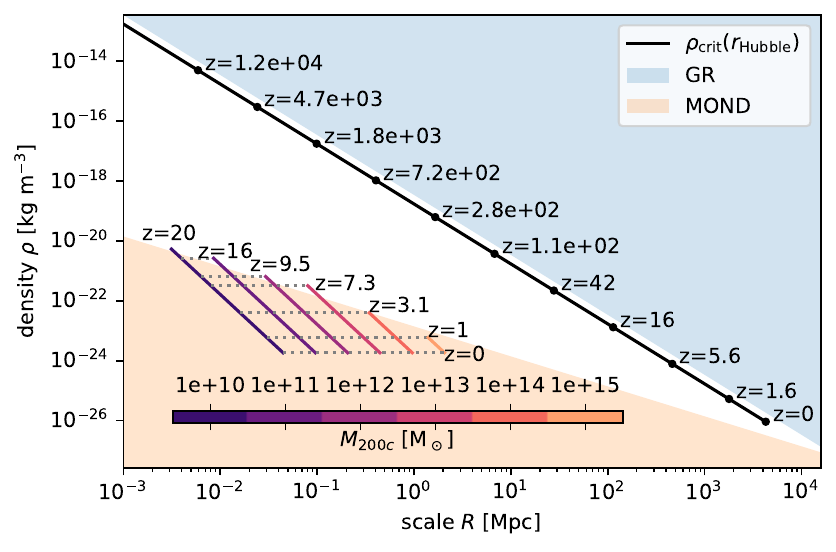}
  \caption{Comparison of the different gravitational regimes as function of scale $R$ and density $\rho$. At each scale, very low-density regions fall in the MOND regime (orange area), while high-density regions fall into the General Relativity (GR) regime (blue area). As a black line, we plot the critical density of the Universe as a function of the Hubble radius, showing the typical regime of the observable Universe. Color-coded the regime for halos of different masses, defined as spherical over-densities with a density contrast of 200 ($M_\text{200c}\approx 10^{10} M_{\odot}$). These lines extend from their formation redshift to the present day.}
  \label{fig:grandmond}
\end{figure}

\subsection{A MONDian cosmological model}
The basic premise of MOND is that in the regime of weak gravitational acceleration, gravity becomes stronger. This transition occurs when the gravitational acceleration $a_\text{gr}<a_0 = 1.2\times 10^{-10}$ m s$^{-2}$. The lack of an existing underlying theory of MOND \citep{1999PhLA..253..273M, 2024arXiv240617930M} and a covariant version of it \citep{2024arXiv240406584B} still puzzles the current status of the theory. As we show in the following, this is not a problem when establishing a MONDian cosmological model.

Consider a region of scale $R$ with mean density $\rho$. The typical gravitational acceleration at the edge of this sphere is given by:
\begin{equation}
    a_\text{gr} = \frac{4 \pi}{3} G \rho R,
\end{equation}
where $G$ is the gravitational constant. Equating this to the transition acceleration from MOND to Newtonian physics, $a_0$, we find the limiting density as a function of scale:
\begin{equation}
    \rho_\text{MOND}\left(R\right) = \frac{3 a_0}{4 \pi G R}.
\end{equation}
At each scale, regions with smaller densities are subject to MOND. This regime is highlighted in orange in Fig.~\ref{fig:grandmond}.

Newtonian Gravity results from the weak field limit of General Relativity (GR), approximately when the gravitational field $\varphi_\text{gr}<c^2$, where $c$ is the speed of light in a vacuum. In our simplified setting above, this results in a limiting density as a function of scale:
\begin{equation}
    \rho_\text{GR}\left(R\right) = \frac{3 c^2}{4 \pi G R^2}.
\end{equation}
GR effects become relevant for regions of scale $R$ if they are denser than this threshold, shown in blue in Fig.~\ref{fig:grandmond}. As can be seen, the two regimes would intersect because of their different trends with scale for $R>10^4$ Mpc, which is irrelevant for observational cosmology.

Indeed, we can plot the critical density of the Universe:
\begin{equation}
    \rho_\text{crit} = \frac{3 H\left(z\right)^2}{8 \pi G},
\end{equation} 
where $H\left(z\right)$ is the expansion rate at redshift $z$. The size of the observable Universe can be approximated by: 
\begin{equation}
    R_H = \frac{c}{H\left(z\right)},
\end{equation}
and so we can plot the typical gravitational regime of the observable Universe for different redshift, shown as a black line in Fig.~\ref{fig:grandmond}. We find that for all redshifts, this lays at the edge of the GR region, justifying its use when describing the evolution of the entire observable Universe.\footnote{Incidentally, $\rho_\text{GR}\left(R_H\right) =  \frac {\rho_\text{crit}}{2}$. So, the placement of the line is independent of the present-day expansion rate $H\left(z=0\right)$, which only sets how far it extends to the right.} Thus, by construction, MOND does not affect the description of the distance redshift relation, of the CMB and of the evolution of linear density fluctuations at high redshifts (which are used as initial conditions for cosmological simulations). 

We also consider halos (gravitationally bound structures) made up of baryons and sterile neutrinos, but for the sake of argument, we shall assume that they are spherical over-densities of 200 times the critical density. This fixes their typical density while their scale is concurrently given by:
\begin{equation}
    R_\text{halo} = \left( \frac{3 M_\text{200c}}{4 \pi 200 \rho_\text{crit} }\right)^{\frac{1}{3}}.
\end{equation}
We plot the resulting scale density regime from the redshift at which these objects could theoretically first be observed in an all-sky survey to the present as colored lines for halo masses ranging from dwarf galaxies $M_\text{200c}\approx 10^{10} M_{\odot}$ to galaxy cluster masses $M_\text{200c}\approx 10^{15} M_{\odot}$ as colored lines in Fig.~\ref{fig:grandmond}. These calculations are all based on the well-studied case of halo formation in the $\Lambda$CDM model. Nonetheless, they fall entirely into the MOND regime. A cosmological investigation based on MOND will thus have to re-establish the fundamentals of halo formation and structure through dedicated simulations, as none of the $\Lambda$CDM simulation-based results are applicable. Luckily, the initial conditions of these simulations can be seeded using GR-based fits to the CMB power spectrum that are unaffected by MOND.

\subsection{Sterile neutrino dark matter}
Originally, the $\nu$HDM model appeared reasonably promising, considering the uncanny successes of MOND on the galaxy scales, such as the baryonic Tully-Fisher relation for spirals, the Faber-Jackson for ellipticals \citep{2012LRR....15...10F} or the Radial Acceleration Relation \citep{2018A&A...615A...3L} (RAR). Early work by \citet{1994A&A...284L..31S} suggested that clusters might lie on the high-mass extension of the Faber-Jackson relation of elliptical galaxies. But clusters proved to be deeply unsettling for MOND. Subsequent efforts by \citet{1999ApJ...512L..23S} revealed a mass discrepancy in gas-rich, X-ray emitting galaxy clusters.  The difference in the ratio of the dynamical to observable virial mass, although is reduced in the context of MOND, a factor of 2
remains present. In the following considerations \citep{2003MNRAS.342..901S}, the non-zero rest-mass (but not yet proposed as sterile) neutrino premise was introduced when attempting to bridge that discordance between MOND theoretical predictions and observations. The first conjecture of the neutrino rest-mass was $m_{\nu_{s}} = 2.2 $ eV/c\textsuperscript{2} ($\Omega_{\nu} \approx$ 0.13 for h = 0.7 which is the unitless parametrization of the Hubble constant like $H_0$/(100 km/s/Mpc)). The \citet{PhysRevLett.42.407} limit would be sufficiently low in order not to affect the galaxy dynamics (leaving them purely baryonic and MONDian), while their free-streaming length would be short enough to not discernibly affect the CMB.

%Note on TremaineGunn limit. If lepton mass <1 MeV and there are some other stable neutral leptons, arguments are valid.
%if If lepton mass > 1MeV then the leptons are relativistic when they decouple thus it is no longer valid NOT possible;

This suggestion was further developed by \citet{2009MNRAS.394..527A} who constrained the sterile neutrino rest-mass 2.2eV/c\textsuperscript{2} < 8eV/c\textsuperscript{2}  $ \lesssim m_{\nu_s} $ < 18eV/c\textsuperscript{2} < 1MeV/c\textsuperscript{2}, after rejecting the hypothesis of an active neutrino, based on their potential but unseen contribution to the total mass-energy budget, or two massive sterile ones (failing to fit the 3rd acoustic CMB peak). \citet{2009MNRAS.394..527A} used CAMB \citep{Lewis_2000} to fit the WMAP5 (Wilkinson Microwave Anisotropy Probe) \citep{2009ApJS..180..306D} observations of the CMB, having varied the following parameters: $\Omega_{\text{b}}$, $\Omega_{\nu_{s}}$, $n_s$, d$n_s\text{/}d\ln k$, $\tau$ and $H_0$. Likewise, we conducted a similar analysis with Bayesian techniques but using the \textit{Planck} \citep{2020A&A...641A...6P} data instead, fitting for $\Omega_{\nu}$, $A_{\text{Pl}}$, $A_\text{s}$, $n_s$, $\tau$, $H_0$ in order to find the optimized $\nu$HDM parameters (hereafter opt-$\nu$HDM).

%\ns{The $H_0$ is the Hubble constant quantifying the expansion rate of the Universe, measured in km/s/Mpc, occasionally treated as $h$, which is its unitless parametrization $H_0 = 100 \times h \text{ km}  \text{/s}\text{/Mpc}$. \linebreak \\}

The energy density of the massive sterile neutrinos $\Omega_{\nu_s}$ is defined in \citet{2020A&A...641A...6P} (their section 7.5.3 - \textit{Joint constraints on neutrino mass and $N_{\text{eff}}$}) as: 
\begin{equation}
    m^{\text{eff}}_{\nu_s} = \Omega_{\nu_s}h^2 \times \left( 94.1 \text{ eV}\right).
\end{equation} 
In general, $\Omega_{\nu}$ includes the contribution of both the active and the sterile neutrinos ($\Omega_{\nu} = \Omega_{\nu_a}\text{ + }\Omega_{\nu_s}$). The physical mass of the sterile neutrino (which is supposedly a thermal relic) is defined as:
\begin{equation}
    m^{\text{thermal}}_{\nu_s} = \left( \Delta N_{\text{eff}} \right)^{-\frac{3}{4}} \times m^{\text{eff}}_{\nu_s}.
\end{equation}
The $N_{\text{eff}}$ is the effective energy density number of the total (active and sterile) neutrinos, and $\Delta N_{\text{eff}}$ is the difference between active and sterile contributions. It induces changes in the freeze-out temperature $T^{\text{FO}}$ of weak interactions $\Gamma_{n\leftrightarrow p} \sim H$ (the interaction rate per unit target particle density $\Gamma$ compared to the expansion rate of the Universe/dilution of primordial hot plasma $H$ \footnote{We specifically use "$\sim$" which stands for "same order of magnitude".}). The higher the $N_{\text{eff}}$, the bigger becomes the expansion rate, and the higher $T^{\text{FO}}$, thus influencing the CMB power spectrum from its 3rd peak to the damping tail (since its 1st peak is determined by the geometry and the 2nd by the baryonic content).

The optical depth $\tau$ at the reionization era is defined as:
\begin{equation}
    \tau \left( t \right) = \sigma_\text{T} \left( t \right) \int^{t_{0}}_{t} n_e \left( t \right) c dt,
\end{equation}
where $n_e$ is the electron number density, $\sigma_\text{T}$ is the Thomson scattering cross-section and $t_0$ is the present time. According to the $\Lambda$CDM scenario, the intergalactic medium was neutral after recombination (z$<$1100), until the re-ionization epoch, which follows at z $\approx$ 6-8.
%Due to any (yet) formed emitting source (either stars or galaxies), and since the Universe was transparent at that particular state, the era was called "the Dark Ages". Thereafter, this era will come to an end by the re-ionization epoch, which follows at z $\approx$ 6-8. How exactly the Universe came over a new ionized condition remains under investigation. 
The re-ionization process affects the formation of stars and galaxies, because these objects are sources of ionized photons. Subsequently, the re-ionization procedure touches the CMB radiation through the Thomson scattering \citep{1999MNRAS.308..854G}.

The parameter $A_\text{s}$ is the initial super-horizon amplitude of curvature perturbations at the pivot scale $k_0 = 0.05 \text{ Mpc}^{−1}$ (the wavenumber at which the primordial power is best constrained):
\begin{equation} \label{pkformula}
    \mathcal{P}_R\left( k\right) = A_\text{s} \left( \frac{k}{k_0}\right) ^{\left[ n_s - 1 \text{ + } \frac{1}{2} \left( \frac{dn_s}{d\ln k}\right) \ln \left( \frac{k}{k_0} \right) \right] } ,
\end{equation}
where $n_s$ is the scalar index for inflation, $dn_s\text{/}d \ln k$ is the running of the spectral index, which is taken as constant. Purely adiabatic scalar perturbations at very early times is assumed, as in \citet{2014A&A...571A..16P}. The observed amplitude scales through the relation $A_\text{s}e^{-2 \tau} \approx \left( 1.884 \pm 0.012 \right) \times10^{-9}$ , where $A_\text{s} \approx 2.101^{\text{+}0.031}_{-0.034} \times 10^{-9}$ at 68\% C.L. according to \citet[their table 2.]{2020A&A...641A...6P}.
%Since the CMB fluctuations are linear up to lensing corrections, and the lensing corrections are largely oscillatory, the average observed CMB power spectrum amplitude scales nearly proportionally with the primordial comoving curvature power spectrum amplitude $A_\text{s}$. The sub-horizon CMB anisotropies are, however, (Thompson) scattered by free electrons that are present after reionization, so 

The $A_{\text{Planck}} = A_{\text{Pl}}$ parameter incorporates the main instrumental effects, coming from the calibration uncertainties of each frequency in temperature and polarization maps. The \textit{Planck} calibration parameter has a prior $A_{\text{Pl}} = \mathcal{N} \left( 1.0000, 0.0025 \right)$. We have used the default namelist CosmoSIS file values (called Planck lite priors). This quantifies the "cleaning" of the sky maps from foregrounds. For large scales ($\ell<30$), there are three polarized components, the CMB, the synchrotron emission and thermal dust emission. For small scales ($\ell>30$), emission from our own Galaxy is masked in different frequencies (100, 143, and 217 GHz). Last, there are point-sources and large objects (Large and Small Magellanic Cloud, the Coma cluster and M31), which are masked with specific techniques.

The $A_{\text{Pl}}$ should not be confused with $A_{\text{Lens}}$. The $A_{\text{Lens}} = A_{\text{L}}$ parameter refers to the CMB lensing, such that $A_{\text{Lens}} = $ 0 corresponds to unlensed, while $A_{\text{Lens}} = $ 1 is the expected lensed result in the standard $\Lambda$CDM model. The lensing depends on the actual amplitude of the matter fluctuations along the line of sight. Its scaling with the gravitational potential becomes visible at large scales, or higher $\ell>3000$ values in the CMB damping tail. Mathematically, it multiplies the matter power spectrum:
\begin{equation}
    C^{\Psi}_{\ell} \rightarrow A_{\text{Lens}} C^{\Psi}_{\ell},
\end{equation}
acting effectively as a fudge lensing factor \citep{PhysRevD.77.123531}, smoothing the CMB peaks.
The $A_{\text{Lens}}$ parameter is of particular interest in this study, because of the small mismatch at $\ell \approx$ 1100 from the \textit{Planck} estimation in Fig. \ref{cmb}, but it is not treated as a free parameter with a prior distribution in our study.

Numerical $\nu$HDM simulations have been pushed firstly by \citet{2011MNRAS.417..941A} who built up a calculation technique to integrate particles’ positions and velocities in a grid following one of the MOND formulations, QUMOND \citep{2010MNRAS.403..886M}. Their cosmological computations reproduced clusters of baryons and neutrinos and MOND galaxies, although low-mass clusters were too few and high-mass ones were too many compared to observations \citep{Rines_2013}.

Secondly, \citet{Katz_2013} demonstrated through N-body simulations that \textit{velocities in the MOND simulations as well as the masses of the largest clusters tend to be higher than those in a comparison $\Lambda$CDM simulation}, trying to explain large cluster bulk flows \citep{Kashlinsky_2008, 2021A&A...649A.151M} and the existence of Bullet-like clusters \citep{Menanteau_2012, 2023ApJ...954..162A}. \citet{Katz_2013}, also previously \citet{10.1111/j.1365-2966.2008.13961.x},  worked with vast box sizes in their simulations (512Mpc/h and 32Mpc/h sided), but they had treated baryons as dissipation-less dust, thus not examining any hydro-dynamical effects. % or sub-grid physics. 

The concern about the formation of the "correct" number of high-mass ($M_{200}>10^{15.1} M_{\odot}$) clusters was exemplified by \citet{10.1093/mnras/stt1564}. These authors have confirmed that this N-body model was overproducing clusters, no matter what the neutrino mass could be in the range 11–300 eV. The hydro-dynamical version of it by \citet{Witt} corroborated it, leaving the idea of a MONDian cosmology in terms of the $\nu$HDM framework unsatisfactory.

From the point of view of Particle Physics, a light, neutral, sterile (right-handed) neutrino is highly motivated as ordinary neutrinos interchange their masses while they propagate. Since all the other fermions (assuming their chemical potential $\mu_{\nu_{s}}$= 0) have, in the standard model, both left- and right-handed components, it seemed tempting to speculate its existence. It is a hot thermal relic \citep{PhysRevLett.103.171301} that never experienced any Boltzmann suppression (exponential decay as the Universe expands, in order for their number density to remain in equilibrium). The dominant production mechanism of a right-handed sterile neutrino is its oscillations at early times \citep{Langacker:1988fp}. It remained relativistic until recombination but was not reheated after decoupling (T$\sim$1MeV). Since it is hot, it has erased all seeds of structures below cluster scale, and its free-streaming length is $\lambda_{\text{fs}} \approx$ 3.5Mpc \citep{10.1093/mnras/staa2348}. In a HDM Universe, the first structures will be pancake-type, while smaller groups or galaxies would form from fragmentation in a so-called Top-down scenario \citep{PhysRevLett.72.17}. Note that a sterile neutrino is also considered a DM candidate with a rest-mass $m_{\nu} \approx$ 7 keV/c\textsuperscript{2} \citep{Bulbul_2014, 2017arXiv170208430M}.

Nonetheless, experiments like Karlsruhe Tritium Neutrino (KATRIN) \citep{PhysRevLett.123.221802} and STEREO \citep{Almazán2023} have set some upper limits for the active neutrino mass and subsequently for the sterile one, but their sensitivity vanishes $\sim$ 10 eV, just around the mass range where our sterile neutrino hypothetical $m_{\nu_{s}}$ falls between. For higher masses/resolution, one has to anticipate their final results, while the latest ones are published in \citet{PhysRevD.105.072004}. The possibility, though, that an ordinary neutrino is a viable DM candidate in the keV range has been excluded by \citet{Aker2022}. However, these experiments do not directly calculate the absolute mass of the sterile neutrino $m_{\nu_{s}}$, but rather the difference ${\Delta m}^2 = m_{\nu_{s}}^2 - m_{\text{active}}^2$. Furthermore, what KATRIN for example measures, is the tritium beta spectrum directly. Tritium $\leftindex^3 {H}$ is a radioactive isotope of hydrogen with the same number of protons and electrons and 2 neutrons. It is unstable and decaying emitting beta-radiation. The measurement of the $\leftindex^3 {H}$ beta spectrum is \textit{independent of any cosmological model and does not rely on assumptions whether the neutrino is a Dirac or Majorana particle}. The $\leftindex^3 {H}$ beta spectrum is the number of beta electron counts at the given energies. In order to interpret the shape of the spectrum, several parameters are to be fitted, like $m^2_{\nu}$. This is the effective anti-neutrino mass (antineutrino $\overline{\nu}$, because tritium decays by beta minus), from the single-$\beta$ decay of molecular tritium, defined as:
\begin{equation}
     m^2_{\nu} = \Sigma^{3}_{i=1} |U_{ei}|^2 \times m^2_i  \quad \left[eV^2\right] ,
\end{equation}
which is an algebraic function of the 3 elements. The $U_{ei}$ are the neutrino matrix elements of the Pontecorvo–Maki–Nakagawa–Sakata matrix \citep{osti_4349231, 10.1143/PTP.28.870} that describes the mixing of neutrino states (amplitudes of mass eigenstates) and the 3 neutrino masses. This is a model of neutrino oscillations, when they propagate freely and interact via the weak nuclear force. The last KATRIN constraint on $m^2_{\nu}$ amounts to ${-0.14}^{\text{+}0.13}_{-0.15}$ $\text{eV}^2$\citep{2024arXiv240613516A}. The uncertainties are at the confidence level of 68\%. As a note of caution, KATRIN uses $m_{\nu}$ as the effective electron anti-neutrino mass and not the sterile neutrino mass that we do.

\section{Methods} \label{cosmosis}
\subsection{Software and techniques}
In order to fit the ESA's \textit{Planck} \citep{2020A&A...641A...6P} CMB power spectrum, we have utilized the CosmoSIS (COSMOlogical Survey Inference System) cosmological parameter estimation code. This is a framework for building inference pipelines, particularly focused on cosmology analysis, with an associated standard library of cosmology components. CosmoSIS uses a Likelihood Function, calculating the probability of the observed \textit{Planck} 2020 data, given the values of the parameters of the $\nu$HDM cosmological model we have chosen. We first computed the Likelihood of the $\nu$HDM model and compared it with that of $\Lambda$CDM. We have used the \citet{Witt} parameters (Table \ref{Likelihoods} and Table \ref{cosmo_params}). For this purpose, we simply exploited the \textit{test} sampler. The test sampler only computes a Likelihood (runs the pipeline) for a single set of values (the original ones from \citet{Witt}). As one can see, this configuration has a significantly smaller Likelihood compared to $\Lambda$CDM ($- \log\mathcal{L} = -5325<  -826 $), inducing a serious lack of confidence in the corresponding model. For the sake of the exercise, we use the \textit{Planck} parameters from table 6 of \citet{2020A&A...641A...1P}, which are the $\Lambda$CDM model best-fit values of the combination of data from \textit{Planck} CMB temperature and polarization power spectra (including lensing reconstruction) without the BAO data.

\begin{figure} 
    \includegraphics[width=0.5\textwidth]{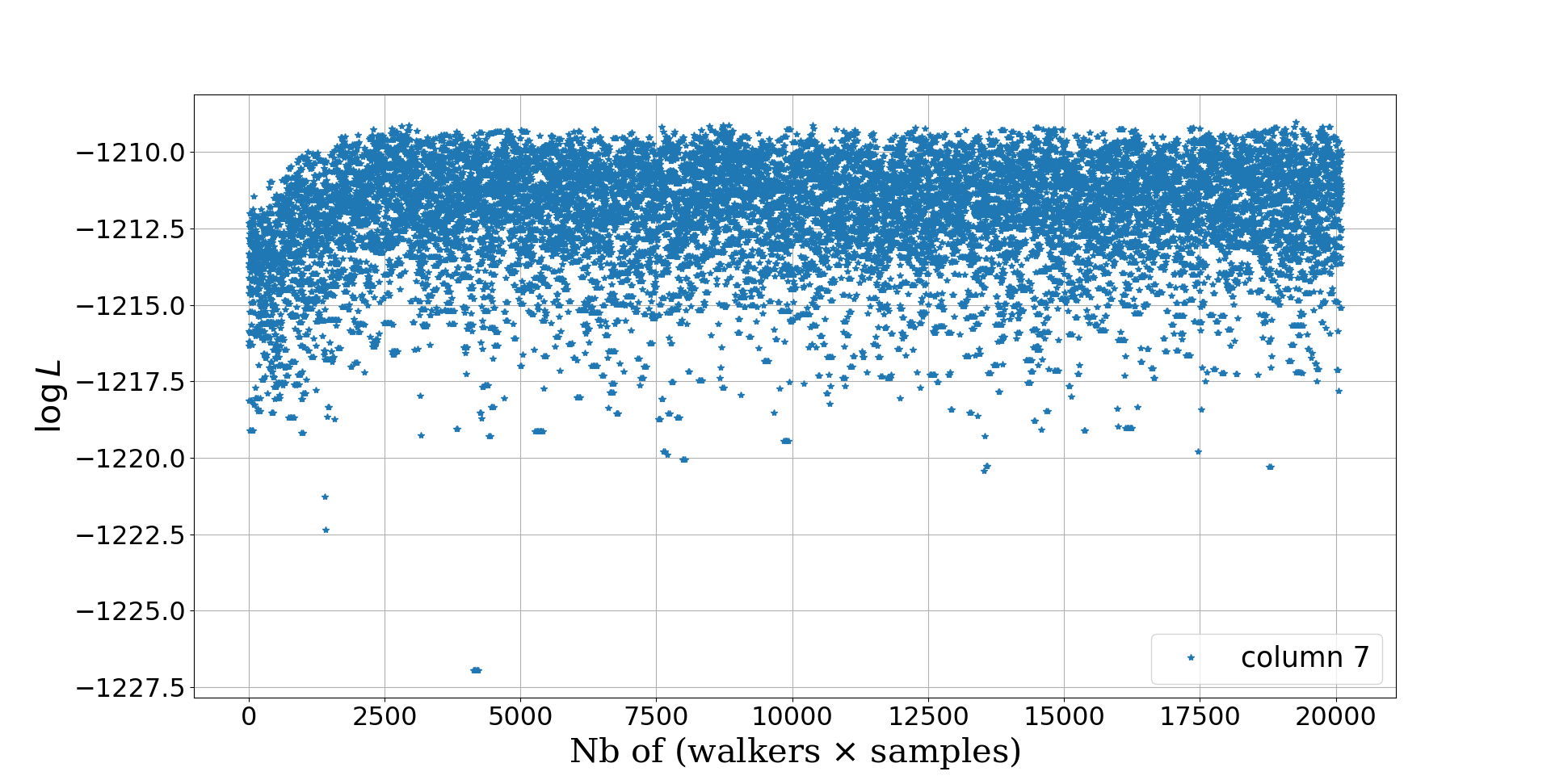} 
    \caption{Likelihood convergence plot from CosmoSIS.}
    \label{Likeli}
\end{figure}

\begin{table}
\centering
\begin{tabular}{llll}
    Model          & $\Lambda$CDM   & $\nu$HDM & opt-$\nu$HDM \\ \hline
    Likelihood     &  -826          &  -5325   & -1210 
    \end{tabular}
    \caption{Likelihoods of the \textit{Planck} data at the best fitting parameters of each model.}
    \label{Likelihoods}
\end{table}

\begin{table} 
\centering
\begin{tabular}{llll}
    Model                 & $\Lambda$CDM         & $\nu$HDM      & opt-$\nu$HDM \\ \hline
    $\omega_{\text{$\nu$}} = \Omega_{\text{$\nu$}}h^2 $      &  0.00065 \textcolor{red} & 0.1205        & $0.1307 \pm 0.001$ \\
    $\omega_{\text{c}} = \Omega_{\text{c}}h^2$   & $0.1200 \pm 0.0012$       &  0.000645     & 0.00001 (F) \\
    $\omega_{\text{b}} = \Omega_{\text{b}}h^2$   & $0.02237 \pm 0.00015$     &  0.022383     & 0.022383 (F) \\
    $H_0$ [km/s/Mpc]      & $67.36 \pm 0.54$     &  67.4         &    $55.64 \pm 0.32$     \\
    $A_{\text{Pl}}$   & $0.997 \pm 0.031$    &  1.0          &    $1.0\pm 0.0025 $    \\
    $n_s$                 & $0.9649 \pm 0.0042$  &  0.966        &    $0.872 \pm  0.0037$    \\
    $\tau$                & $0.0544 \pm 0.0073$  &  0.0543       &    $0.0375\pm 0.0069$      \\
    $\ln(10^{10}A_\text{s})$     & $3.044 \pm 0.014$    &  3.0448       &    $2.987\pm 0.015$      \\
    $\Omega_{m}$          & $0.3153 \pm 0.0073$  &  0.3          &    $0.495 \pm 0.009$
    \end{tabular}
    \caption{Cosmological Parameters for the 3 studied models. For $\Lambda$CDM, these are the best fits from \citet[their table 7]{2020A&A...641A...1P}. For $\nu$HDM, the model parameters are the \citet{Witt} values.
    For the opt-$\nu$HDM, they are inferred from \href{https://cosmosis.readthedocs.io/en/latest/reference/samplers/emcee.html}{CosmoSIS}. The theoretical model parameters $\omega_\text{c}$ and $\omega_{\text{b}}$ are kept fixed (F).} 
    \label{cosmo_params}
\end{table}

Secondly, we built another Likelihood function through the \textit{emcee} module on our opt-$\nu$HDM model. The input values are from  \citet{2020A&A...641A...5P}. The actual configuration was 20 walkers (points exploring the parameter space). It tells CosmoSIS to generate 1000 samples per walker, and to save results to the hard disc every 15 steps. We have reached the convergence plot by producing a chain of values (list of vectors of the parameters). When we achieve the stationary state, the evaluation of each of the parameters is done by looking at the trace plots, shown for likelihood evaluation of each walker in Fig.~\ref{Likeli}. After an initial burn-in phase, which we cut off in subsequent analysis steps, the walkers clearly reach a steady state.

More particularly, the CosmoSIS module \textbf{mgcamb} \footnote{\url{https://cosmosis.readthedocs.io/en/latest/reference/standard_library/mgcamb.html}} is used. This is a modified version of CAMB, in which the linearized Einstein equations of GR are modified so that the sterile neutrino is incorporated. Note that the CDM component cannot be set exactly to 0, that's why there is a very minor contribution still active (See Table. \ref{cosmo_params}). Also, there is no one-to-one equivalence between \textbf{mgcamb} of CosmoSIS and CAMB, although the former makes use of the latter. This might create complications if one wants to translate each of the parameters of the one program to another. CAMB is capable of treating information on the number of different eigenstates of the massive neutrino, while \textbf{mgcamb} cannot. 

As in CAMB, so in CosmoSIS, we implement the existence of the massive sterile neutrinos (\texttt{massive\_nu =  1}, sterile\_mass\_fraction = 1) through the software setup parametrization, modifying the namelist file. The parameter $N^{\nu\text{HDM}}_{\text{eff}}$ = 4.048 accounts for the active (massless\_nu = 2.0293) and the sterile massive part. The standard \textit{Planck} value in $\Lambda$CDM is $N^{\Lambda\text{CDM}}_{\text{eff}}$ = 3.046, by assuming the existence of three active neutrinos \citep{MANGANO20028}. 

\begin{comment}
%Referee said " translating the variable names of physical quantities in a computer code into the more standard names should not be a task of the reader" 
The CosmoSIS.ini namelist file contains: \newline
\texttt{
massless\_nu = 2.0293 \newline
massive\_nu =  1 \newline
sterile\_neutrino = 1 \newline
sterile\_mass\_fraction = 1 \newline
standard\_neutrino\_neff = 4.048 \newline
delta\_ neff = 1}  
%maybe WRONG definition $\ln \left( 10^{10}A_\text{s} \right) = N_{\text{eff}}$
\end{comment}

Lastly, CosmoSIS needs a file as \textbf{Priors} which is not modified, since we are comparing the \textit{Planck} data after removing all the background noise from our own galaxy, extra-galactic sources and the Sunyaev-Zeldovich (SZ) effect.

The program makes use of Bayesian statistics. In order to constrain a cosmological model M of astrophysical data D, a posterior distribution p($\bm\theta|D, M$) is calculated on the space of cosmological parameters $\bm{\theta}$, using Bayes' theorem: 
\begin{equation}
    p\left( \bm\theta|D,M \right)= \frac{L \left( D|\theta,M \right)}{E \left( D|M \right)}p\left( \bm \theta \right),
\end{equation}
where $p \left( \bm \theta \right)$ is the prior, $L\left(D|\bm \theta, M \right)$ is the Likelihood and $E\left(D|M\right)$ is the evidence. In this task, M is the $\nu$HDM model and D is the \citet{2020A&A...641A...6P} data.
%To determine whether a given data set, D (Planck), prefers a model $M_1$ ($\nu$HDM) or model $M_2$ (opt-$\nu$HDM), we adopt the deviance information criterion (DIC). Considering the generalized chi-squared $\chi^2 \left( \bm \theta \right) = -2 \ln L \left( D | \bm \theta , M_{i} \right)$, the mean goodness of fit over the posterior volume is evaluated as $\langle\chi^2 \rangle= -2 \langle \ln L(D|\bm \theta, M_i) \rangle$. A model which fits the data better will have a lower $\langle\chi^2\rangle$.
\subsection{CosmoSIS Priors}
\begin{table} 
    \centering
    \begin{tabular}{cc}
        Parameter  & Prior range \\
        \hline
        $\omega_{\nu}$ & $\mathcal{U}$(0.1, 0.15)\\
        h  & $\mathcal{U}$(0.3, 0.8)\\
        $\tau$& $\mathcal{U}$(0.01, 0.075)\\
        $n_s$  & $\mathcal{U}$(0.7, 1.0)\\
        $\ln \big( 10^{10}A_{s} \big)$ & $\mathcal{U}$(2.0, 3.1)\\
        $A_{\text{Pl}}$ & $\mathcal{N}$ (1, 0.0025)\\
        $\Omega_{\text{b}}$ & Fixed to 0.0223828\\
        $\Omega_{\text{k}}$& Fixed to 0\\
    \end{tabular}
    \caption{Priors on the unitless parameters used in the fitting of the  \citet{2020A&A...641A...5P} values. $\mathcal{U}$(min,max) stands for a uniform distribution between "min" and "max". $\mathcal{N} \left( \mu, \sigma \right)$ stands for a normal (Gaussian) distribution centered on mean $\mu$ and with standard deviation $\sigma$.}
\end{table}
We have decided to perform the task of fitting the $\nu$HDM parameters with the same (6) number of parameters as in $\Lambda$CDM model. The choice of the parameters is based on the characteristics of the theoretical model ($\Omega_{\nu}$ in $\nu$HDM, instead of $\Omega_{\text{CDM}}$ in $\Lambda$CDM), while $A_{\text{Pl}}$, $\ln \left( 10^{10}A_\text{s}\right)$, $n_s$, $\tau$ and $h$ are treated as free in both the $\Lambda$CDM and the $\nu$HDM model. If one imposes more free parameters, the fit may be even better, but this decreases the model's confidence.

The range of the priors has been adjusted in the process. For $\omega_{\nu}$, the range has been set to [0.1, 0.15]. On the other hand, $h$ was originally set to a wide interval of [0.5, 0.8], but it was shown, during the MCMC run, that the range required to be enlarged for a more stable value. This demonstrated the unpredictability of a high-dimensional parameter space. The $A_{\text{Pl}}$ was set to its default (usual $\Lambda$CDM) range and it never fluctuated significantly. As for the $A_\text{s}$ value, due to its correlation with $\tau$, it was expected to vary slightly around its initial value.

The degeneracy of $\tau$ and $n_s$ is naturally seen, since these are main players shaping the CMB power spectrum. The optical depth $\tau$ was restricted to an upper value of 0.1, since this was approximately the value ($\tau \approx 0.166$) of the first WMAP results \citep[their table 1]{2003ApJS..148..175S}. Its lower limit $\tau < 0.01$ appeared on the same trend as the subsequent estimated CMB experiments, reporting a decreasing value from the first \textit{Planck} results of $\tau \approx  0.079 \pm 0.017$ \citep[table. 3]{2016A&A...594A..13P} till the latest $\tau \approx  0.0580 \pm 0.0062$  \citep[table. 5]{2024A&A...682A..37T}.

The $n_s$ upper value represents the standard approximation of Harrison-Peebles-Zeldovich \citep{PhysRevD.1.2726, 1970ApJ...162..815P, 10.1093/mnras/160.1.1P}. The lower value was set to 0.7, as a rather opposite expectation of a scale invariance of $n_s$ = 1, supposing that MOND gravity will reinforce the perturbations later as they proceed and grow in the horizon.  By construction, the $\nu$HDM model and its variants have $\Omega_{\text{CDM}}\ll \Omega_{\nu}$. We have fixed this value at $\Omega_{\text{CDM}} h^2 = $ 0.00001, since the program crashes at the exact 0.0 value. The $\Omega_{\text{b}} h^2 = $ 0.0223828 is fixed, like \citet{Witt}, as is the $\Omega_\text{K} = $ 0.0. The helium mass fraction is also set to $Y_{p} = $ 0.25.These values have been kept fixed during this exercise and are noted as "(F)" in Table \ref{cosmo_params}.

It is noteworthy to mention that the Bayesian fitting process is designed to stop after a number of iterations, examine the corresponding convergence and restart with a more optimal parameter range. As a side note, we indicate that CosmoSIS uses the \textit{Planck} data \citep{2020A&A...641A...6P}, but there is an updated version in \citet{2024A&A...682A..37T}. We have affirmed that the $\nu$HDM cosmological model is a conservative idea, meaning that the Big Bang hypothesis and the inflationary epoch were kept in place to provide the seeds for the structures which will be forming thereafter.

%\xrightarrow[\text{vocabulary}]{\text{CosmosSIS}} yhe =
%\xrightarrow[\text{vocabulary}]{\text{CosmosSIS}} \omega_{c} h^2 =

\begin{comment}
\begin{itemize}
    \item A uniform prior in the neutrino energy density:\\$0.10<\omega_{\nu} = \Omega_{\nu} h^2<0.15$.
    \item A uniform prior in the unitless Hubble parameter:\\ $30<h = \frac{H_0}{100} < 80$.
    \item A uniform prior in the reionization optical depth: $0.01 < \tau< 0.07$.
    \item A uniform prior in the scalar spectral index:  $0.7<n_s < 1 $.
    \item A uniform prior in the structure amplitude parameter:\\ $2.0 < \ln \big( 10^{10}A_{s} \big) <3.1$.
    \item A Gaussian prior in the total Planck calibration $A_{\text{Pl}}$ at map level with a mean 1.0 at and standard deviation $\sigma_{A_{\text{Pl}}} = $ 0.0025.
\end{itemize}
\end{comment}

%: $0.9 < A_{\text{Pl}}< 1.1$.

\begin{figure*} 
  \includegraphics[width=\textwidth,height=10cm]{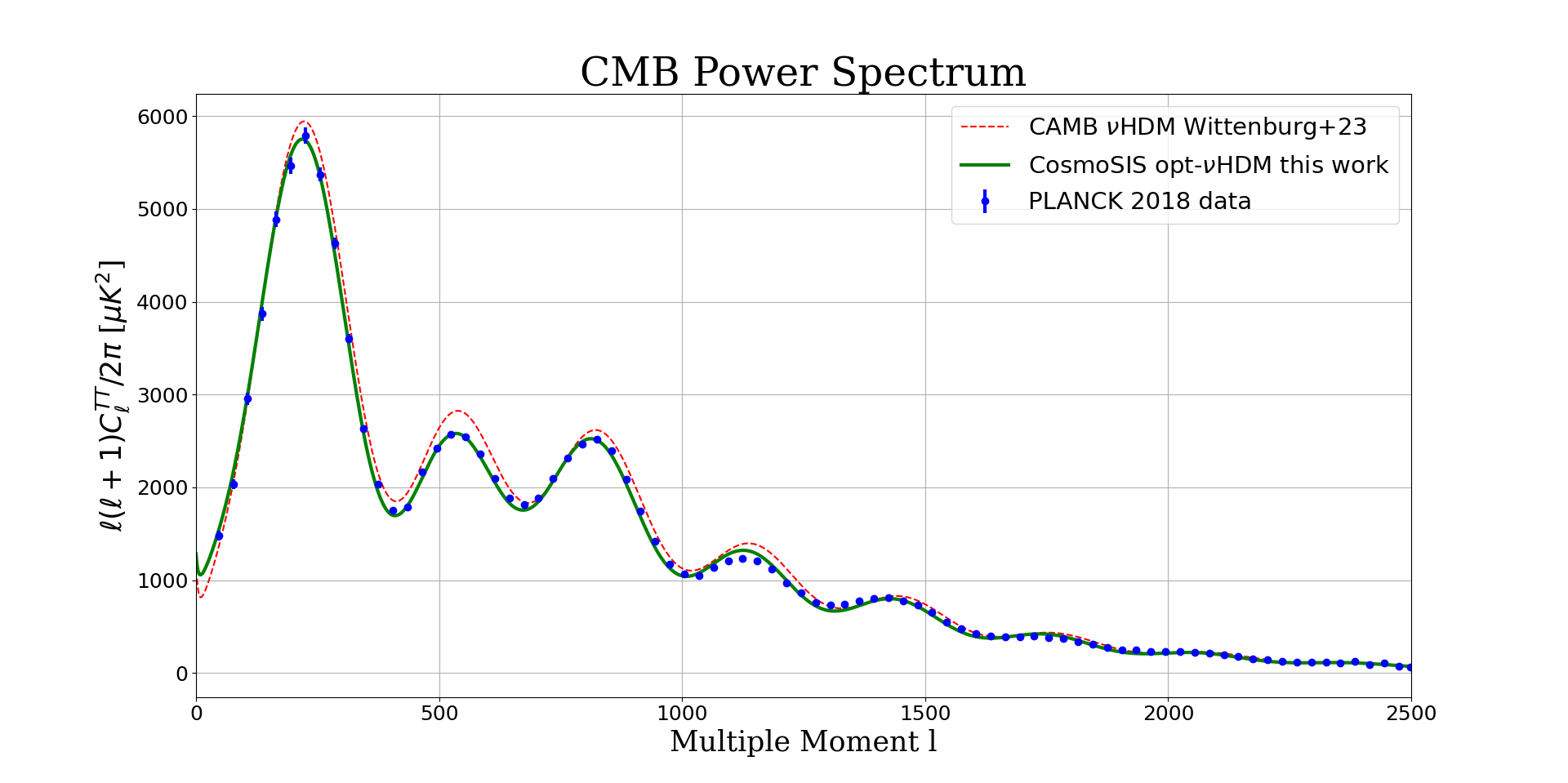}
  \caption{The Temperature Fluctuations power spectrum of the CMB for the 2 different cosmological $\nu$HDM models and the \textit{Planck} 2018 data.}
  \label{cmb}
\end{figure*}

\begin{figure*} 
  \includegraphics[width=\textwidth,height=10cm]{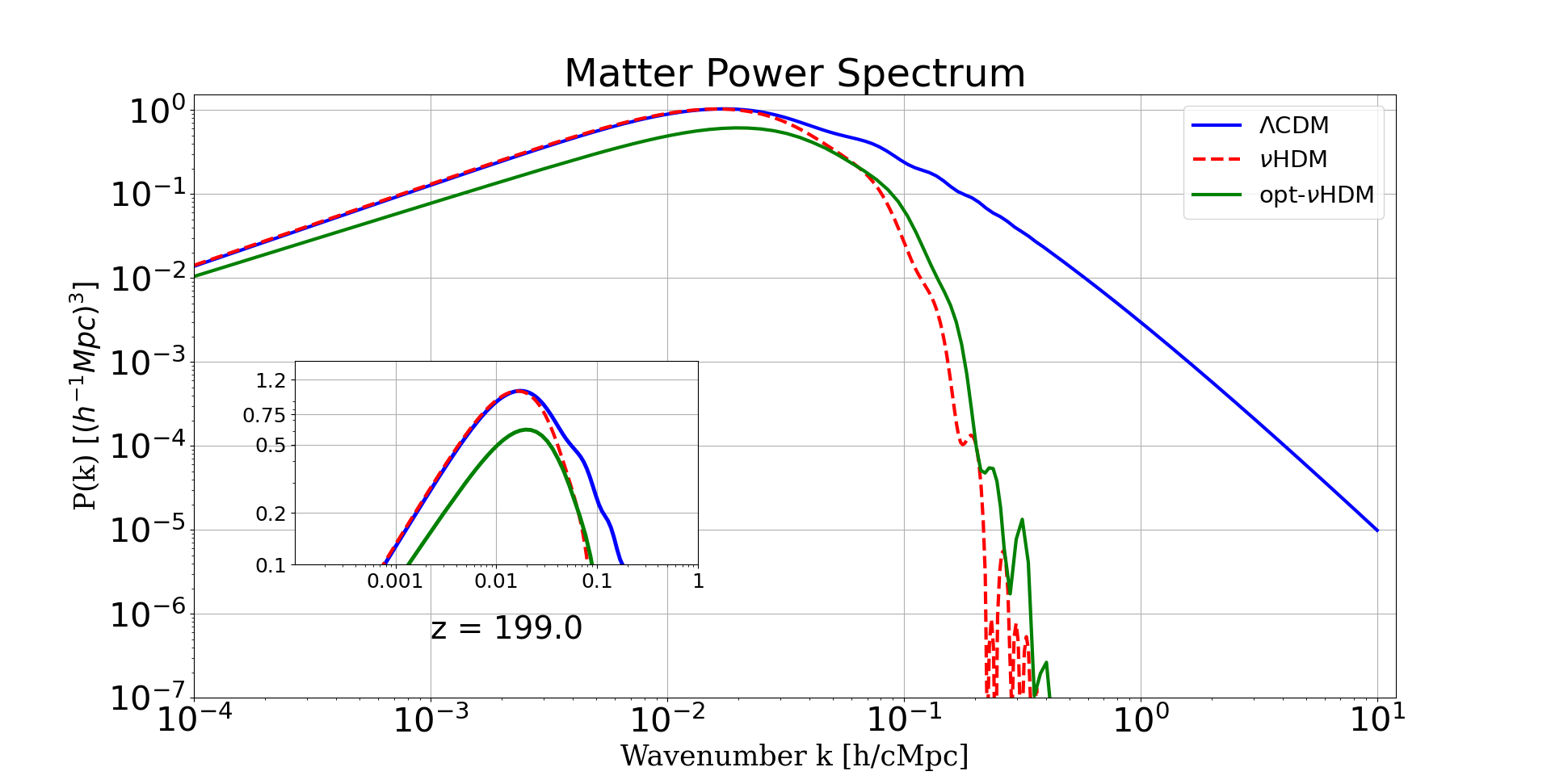}
  \caption{Matter power Spectrum for the 3 different Cosmological Models at z=199.0.}
  \label{pk}
\end{figure*}

\begin{figure*} 
  \includegraphics[width=\textwidth]{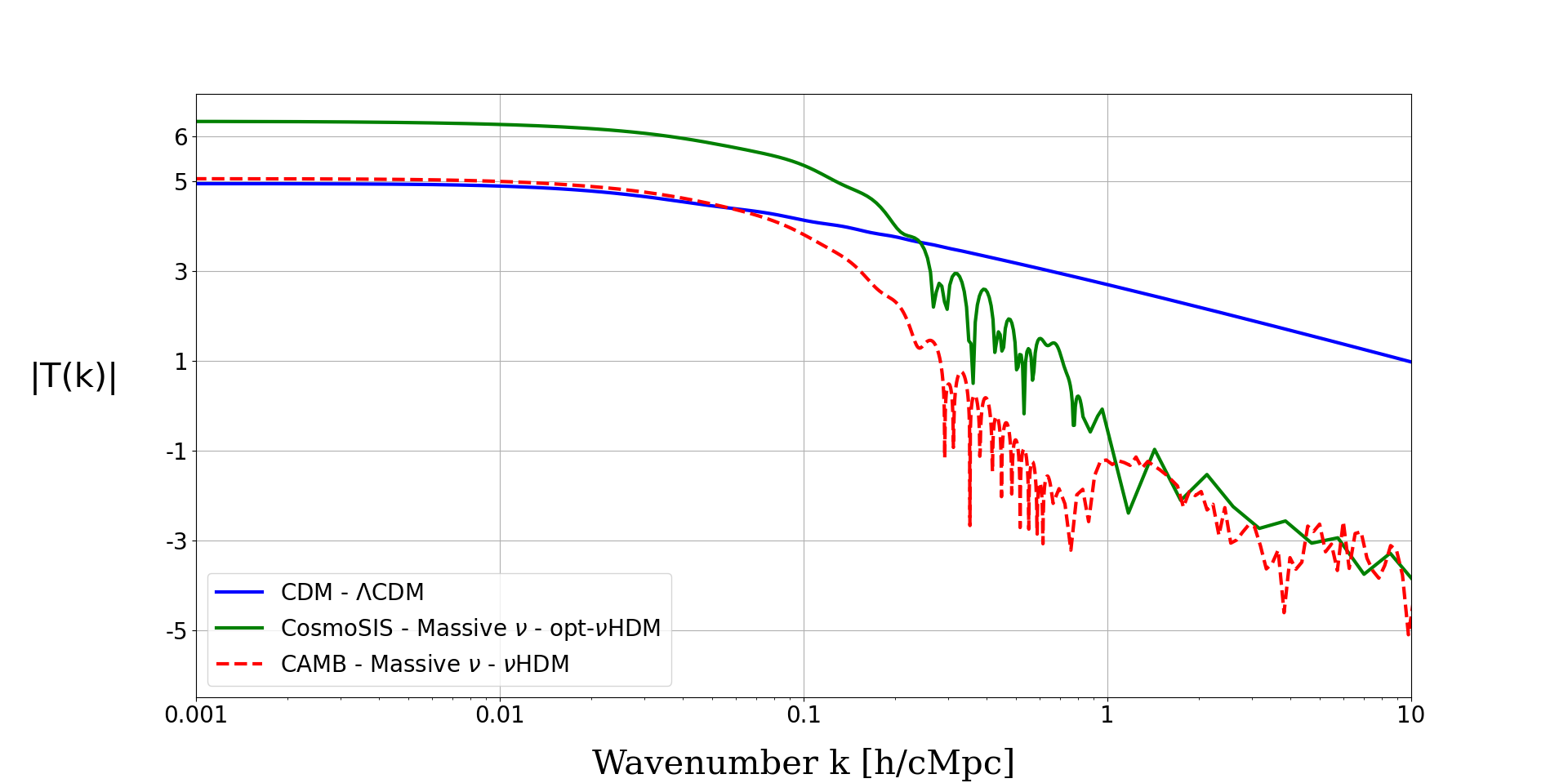}
  \caption{Transfer functions for the 3 different Cosmological Models at z=199.0.}
  \label{transfer}
\end{figure*}

Last, as a technical note, all the pipelines run on a small local machine of 16G of RAM and 8 cores. 

%However, if one wants to do full hydro-dynamical simulations with the opt-$\nu$HDM model parameters, then one would definitely need a parallelized approach on a High Performance computing cluster, which is out-of-the scope of this work, but in preparation. (referee said)

\section{RESULTS} \label{results}
The main motivation of this work was the optimized fitting of the CMB power spectrum (Fig. \ref{cmb}) and the converged outcome values for the model. Before analyzing our results, we underline that we are using the same colors for consistency, red for the $\nu$HDM model, blue for the \textit{Planck} data, which \citet{Witt} call the $\Lambda$CDM model and green for our opt-$\nu$HDM one. It is of course possible to match the data even better if one lets more parameters vary. Nevertheless, one should also keep in mind that a standard $\Lambda$CDM model fit has 6 free parameters: the angular size of the sound horizon at decoupling $\theta_{\text{MC}}$, the cold dark matter $\Omega_{\text{CDM}}h^2$ and baryon density $\Omega_{\text{b}}h^2$, the optical depth at re-ionization $\tau$, the amplitude $A_{\text{s}}$ and the spectral index $n_s$ of inflationary scalar perturbations \citep{2021ApJ...908L...9D}.

Having ensured that the posterior sampling is well converged, we list the mean values of the posterior of the opt-$\nu$HDM model parameters in Table \ref{posterior_mean}. The mean values can vary only slightly from the best-fit ones. The value of $\omega_{\nu_{\text{s}}}$ was increased by 8\%, which will manifest in hydro simulations, providing a most likely different clustering behavior. The shape of the resulting "galaxies" or "clusters of galaxies" needs be investigated. The problem of too many and too massive galaxy clusters reported by \citet{2011MNRAS.417..941A}, \citet{Katz_2013} and \citet{Witt} will be reevaluated upon those calculations. However, the difference in $\omega_{\nu}$ will provide small-scale inhomogeneities that will be unlike those by \citet{Witt}, since the free-streaming length of the neutrino will change, thus their gravitational "grouping" will be happening on larger scales. The extracted matter power spectrum is anticipated to enlighten the physical reason for the mismatch at $\ell \approx 1100$.

One of the striking results obtained by this work is the high value of $\Omega_m$. The quantitative increase, by more than 60\%, gives the flat (no-curvature) opt-$\nu$HDM Universe a rather distinct background evolution. In this case, Dark Energy is dominating marginally over the matter content, strongly deviating from its original construction by \citet{2009MNRAS.394..527A}. In such a heavy Universe, its size will be much smaller compared to the $\Lambda$CDM model, because of the decreased $H_0$ value.

Further on, the reionization epoch happens much later in the opt-$\nu$HDM model. The mean redshift for the reionization takes place much later compared to $\Lambda$CDM, at $z^{\text{opt-}{\nu}\text{HDM}}_{\text{reion}} \approx$ 5.9, which corresponds to $\approx$ 955 Myr after the BB. In a $\Lambda$CDM scenario, it occurs earlier, $z_{\text{reion}} \approx 7.6 \approx$ 700 Myr after the BB. The duration of the opt-$\nu$HDM Dark Ages, up to almost 1 billion years, will probably make its structure formation and evolution very difficult to appear early enough to be consistent to JWST observations. But this will be illuminated by the cosmological simulations. Last, since the matter content is increased, this might also affect the transition to the MONDian regime, that is the time when the perturbations will eventually fall into the low-acceleration regime and MOND will be applied. Note that \citet[their section 2.4]{Witt} are claiming that non-linear MOND effects will be significant at $z\approx50$, but this may change to lower times.

\begin{table} 
    \centering
    \begin{tabular}{ccc}
        Parameter & Posterior Mean value & $\Lambda$CDM\\
        \hline
        $\omega_{\nu} = \Omega_{\nu} h^2$ & 0.1308 & 0.000645\\
        $\Omega_m \left( =  1 -\Omega_{\text{K}}-\Omega_{\Lambda} \right) $ & 0.494306 & 0.309727\\
        $\Omega_{\text{K}} $ & 0.000000 & 0.00000\\
        $\Omega_{\Lambda}$ & 0.505694 & 0.690194\\
        $\omega_{\text{b}} = \Omega_{\text{b}} h^2 $ & 0.022383 & 0.022260 \\
        $\omega_{\text{c}} =\Omega_{\text{c}} h^2 $ & 0.000010 & 0.118800\\
        $H_0$ [km/s/Mpc]     &  55.64        & 67.4    \\
        $z_{\text{reion}}$  & 5.941 & 7.686 \\
        $\tau_{\text{reion}}$ &  0.0375 & 0.0543\\
        100$\theta$ (CosmoMC)  & 1.041467 &1.040777\\
        $m_{\nu_{\text{s}}} $[eV] & 12.202 & 0.060\\
        $\tau_{\text{recomb/Mpc}}$ &  282.71 & 281.32\\
        $\tau_{\text{now/Mpc}}$ &  14263.7 &14193.7\\
    \end{tabular}
    \caption{Mean values for the opt-$\nu$HDM model from the CosmoSIS Posterior distribution compared to $\Lambda$CDM ones from CAMB.}
    \label{posterior_mean}
\end{table}

% NOTE from CosmosSIS : 1.00 nu, $m_{\nu_{\text{s}}}\times c^2/k_B /T_{\nu_{\text{s}}0}  = $ 72630.05

We can use the best-fit values to analyze the goodness of fit of the different cosmological models by predicting the CMB power spectrum and over-plotting it with the data, shown in Fig.~\ref{cmb}.  It is clear from \citet[their left fig. 2]{Witt} that, although the ICs match the observed CMB from \textit{Planck} qualitatively, the matching is not sufficiently good. All the peaks are offset, especially the second one. The exact position and amplitude of the peaks $\left( \ell, \frac{\ell \left( \ell \text{+}1 \right) C^{TT}_{\ell}}{2\pi} \right)$ are in Table \ref{peaks}. Using Bayesian inference tools to actually fit the CMB power spectrum, instead of adjusting parameters ad hoc and by hand, results in a much better fit to the overall CMB power spectrum (Fig. \ref{cmb}). Also, the positions and amplitudes of the peaks are much more accurate (Table \ref{peaks}).
%The angular power spectrum (y-axis component $C^{TT}_{\ell}$) measures the amplitude of the expansion coefficients as a function of the wavelength when they are decomposed in spherical harmonics.

The first peak of the CMB power spectrum is governed by the geometry and the expansion rate of the cosmological model. The opt-$\nu$HDM model is flat (k $= 0$) and the expansion rate is decreased $\approx 18\%$ with respect to $\Lambda$CDM. The second peak is the baryometer, mostly driven by the amount of baryons in the Universe; $\Omega_{\text{b}}h^2$ is fixed in the two $\nu$HDM examined models. The third peak and the damping tail are steered by the effective number of degrees of freedom $N_{\text{eff}}$, which accounts for the contribution of the active and the sterile neutrinos. The small mismatch at the fourth peak (Fig. \ref{cmb}) would possibly be explained from the CMB lensing, but in MOND this is a non-trivial task to do. Latest efforts to tackle this were undertaken by \citet{2022PhRvD.106j4041S}, in the so-called aether scalar tensor (AeST) model.

The power spectrum, depicted in Fig. \ref{pk}, is calculated at redshift $z = 199$ or scale factor $a = 0.05$. Note that the matter power spectrum and the transfer function are plotted against comoving wavenumber (x-axis of Fig. \ref{pk} and Fig. \ref{transfer}). This is a direct comparison with \citet[their fig. 2]{Witt} and it will shift upwards for decreasing redshifts. It demonstrates the larger-scale clustering of the neutrino models, which have essentially no power at galaxy scales ($k\gtrapprox10$ h/Mpc), where massive CDM halos are expected to form in the $\Lambda$CDM model. The neutrino mass difference is indistinguishable in this plot (Fig. \ref{pk}). On large sales $k<10^{-2}$ h/Mpc, structure formation for all the models is linear and driven by Newtonian gravity. At the intermediate scales, $0.1 \text{ h/Mpc} < k < 1$ h/Mpc, the neutrinos become gravitationally bound at the galaxy clusters' scale and effectively show no power at all in galaxies ($1\text{ h/Mpc}<k<10$ h/Mpc), where MOND provides the dynamics.

The opt-$\nu$HDM Universe is almost matter-dominated; the rest-mass of the sterile neutrino component is only increased by 10\%, from 11 eV/c\textsuperscript{2} to 12.2 eV/c\textsuperscript{2} in comparison to the $\nu$HDM model. The baryons are kept in the same percentage; thus, from the resulting $\Omega_{m}\approx$ 0.49, about 0.41 of it will come from neutrinos and the rest, 0.08, will be due to baryons. This means that most of the over-densities will be neutrino-driven, and massive halos of neutrino clouds will surround the baryonic cores. The phenomenon is also portrayed on the transfer function (Fig. \ref{transfer}), at scales of $k<1$ h/Mpc. The 20\% surplus (in log space) of power will have a significant effect on structure formation.  Note that the transfer function is plotted as the absolute value, since any negative sign would apply to under-densities and positive ones to over densities, and both should be accounted for. Last, the transfer function, which describes the evolution of perturbations from the primordial power spectrum (radiation-matter equality epoch) to later matter-dominated era, is calculated with the Newtonian gravitational potential and not with the Milgromian one, since the low accelerations ($a\ll a_0\approx 1.2 \times 10^{-10}$ m/s\textsuperscript{2}) have not yet been reached.
%One can notice that on large scales, $k\approx0.01$ h/Mpc, the opt-$\nu$HDM model is 10\% stronger (in log space) than the Cold Dark Matter component in the standard $\Lambda$CDM.
%This is anticipated because of the different properties of the DM particles; neutrinos follow a top-down formation scenario, also exhibiting neutrino oscillations at smaller scales, which correspond to galaxy cluster scales of 3.5 Mpc.
\section{Discussion} \label{Discussion}
We discuss, in the following section, the implications of the estimated cosmological parameters, the implied initial conditions for hydro-dynamical simulations in the opt-$\nu$HDM with MONDian gravity, and predictions for distance measures.

\subsection{Model assumptions and Cosmological parameters}
Both the $\Lambda$CDM and $\nu$HDM models make use of the same basic assumptions like isotropy, homogeneity, and a metric theory of gravity. The estimated convergence to homogeneity and isotropy takes place at $\approx$ 250 Mpc \citep{2016A&A...594A..16P}. All these hypotheses are encoded in the FLRW metric. Note that this is already implemented in the Phantom of RAMSES code \citep{2015CaJPh..93..232L}, and that's how \citet{Witt} had the same expansion history as $\Lambda$CDM. However, these assumptions cannot account for the magnitude and extension of the observed matter density contrasts, like KBC-type voids \citep{Keenan_2013, 2024MNRAS.527.4388M}. Studies on the galaxy clusters front, like \citet{2021A&A...649A.151M} tackling the isotropy of the local Universe, report an anisotropy of galaxy cluster scaling relations, utilizing observations in X-rays, infrared and submillimeter. \citet{2021A&A...649A.151M} mention that a $\approx 5.5 \sigma$ anisotropy towards a specific direction of the sky $\left(l , b \right) \approx \left( 280^{\circ \text{ +}35^{\circ}}_{ \text{ -}35^\circ}, -15^{\circ \text{ +} 20^{\circ}}_{\text{ -}20^\circ}\right)$ could be a product either of an anisotropic expansion rate $\approx 9\%$ spatial variation of $H_0$, or of an enormous bulk motion of $\approx 900$ km/s, contradicting the $\Lambda$CDM assumptions.

Bulk flows of matter, as reported by \citet{2023MNRAS.524.1885W}, of about 450 km/s within a distance of about 270 Mpc are found to be consistent with the $\Lambda$CDM model with a probability of 2.1$\times 10^{-6}$. These observed bulk flows are accounted for, though in the $\nu$HDM model, and are explained due to gravitationally driven outflows from the observed KBC supervoid \citep{2024MNRAS.527.4388M, 2025MNRAS.536.3232M}.

Additionally, a standard interpretation of the Lilly-Madau plot \citep{1996ApJ...460L...1L, 1996MNRAS.283.1388M} is the quantification of the history of the cosmic star-formation of the Universe by showing the co-moving star-formation rate density over cosmic time. However, \citet{2023MNRAS.524.3252H} derived that the maximum star-formation rate density at "cosmic-noon"  $\left( 2\lesssim z \lesssim 3 \right)$ is due to an over-density at 5 Gpc comoving distance, reconstructing the star-formation histories of the nearby $\approx$ 1000 galaxies. Additionally, 
\citet{2015ApJ...810...47J} and \citet{2017A&A...597A.120J} found that the Southern hemisphere has significantly more early-type galaxies, which appears to align with the same hemisphere having more CMB power.
%\citet{} also investigated the issue of the Northern hemisphere having more power in the CMB dipole than the Southern, hinting that SN data have not been cleaned from all possible systematics.

Last, \citet{Schwarz_2016} and \citet{2023arXiv231012859J} support the violation of statistical isotropy and scale invariance of inflationary perturbations due to the so-called CMB anomalies.  This poses the philosophical question: does fitting the ample amount of extra-galactic data effectively require us to make basic assumptions like isotropy, homogeneity, and a metric theory of gravity? Can observational cosmology even proceed as a field without making these assumptions?

%We emphasize the fact that the products of the Bayesian technique are only relevant for a MONDian $\nu$HDM cosmology and not a standard $\Lambda$CDM model, which faces many issues too \citep{2022PASP..134l1001M, 2023arXiv230911552K}.

Why the MOND phenomenology is arising needs to be investigated. \citet{1999PhLA..253..273M} attempts to explain the appearance of a fundamental acceleration constant $a_0$ as a vacuum effect. Is the $a_0$ varying through the cosmic time? It might be irrelevant to speculate on the spectral index of an exponential abrupt expansion at the very early moments of the Universe or the optical depth of primitive gas on a distinct and dissimilar scheme of MOND-based large-scale gravity, since we are speculating the same initial conditions for models with divergent gravitational behaviors. For MOND gravity, a covariant approach has only recently been possible but does not yet follow the Occam's razor since additional assumptions are required and new aether-type fields are postulated to explain observed physics \citep{PhysRevLett.127.161302}. Thus, it is a slippery task to assume that MOND gravity or a CDM particle component would have the same origin. Nevertheless, \textit{Planck} data should obviously be explained and re-interpreted in any consistent theory.

That being said on the model assumptions, the opt-$\nu$HDM cosmological parameters are now discussed. The parameter whose value deviates the most from the $\Lambda$CDM fit is the Hubble constant $H_0$. We find $H_0 =55.64 \pm 0.32$ km/s/Mpc in the  opt-$\nu$HDM model, significantly smaller than $H_0 = 67.36 \pm 0.54$ km/s/Mpc in the $\Lambda$CDM model. The resulting $H_0$ may seem too small, exacerbating the already existing $H_0$ tension \citep{2023arXiv230110572D}. One could argue that the opt-$\nu$HDM is thus ruled out. 

However, the locally larger value of $H_0 = 73.04 \pm 1.04$ km/s/Mpc by \citet{2022ApJ...934L...7R} may be consistent with a global $H_0 \approx 56$ km/s/Mpc, if the Universe is locally highly anisotropic \citep{2020MNRAS.499.2845H, 2023MNRAS.524.3252H}. The existence of a giant void inflates the redshift gradients and enhances the locally-inferred expansion rate \citep{2024arXiv241000804B}. It would, thus, be premature to disregard the opt-$\nu$HDM model.
 
Moreover, a smaller Hubble constant predicts an older Universe, $t_0 \approx 14.9$ Gyr. This is a rather old age compared to the predictions from the SNe Ia, Cepheids and TRGB (the tip of the red giant branch) data from Pantheon+SH0ES results \citep{2022ApJ...938..113S} of $H_0 \approx72.53 \pm 0.99$ km/s/Mpc, yielding an age estimate of $\approx$ 13.8 Gyr. However, \citet{2023ApJ...953..149C} studied the oldest objects (globular clusters, stars, white dwarfs, and ultra-faint and dwarf spheroidal galaxies) with accurate age estimates to deduce a one-sigma upper limit on the age of the $\Lambda$CDM Universe of $t_0$ = 15.3 Gyr. As these authors mention, their results are sensitive to the priors selections, while the three-sigma upper limit is $t_0$ = 16.5 Gyr and $H_0 \approx$ 55.2 km/s/Mpc. \citet{2024arXiv240718307N} also found that problematic for the $\Lambda$CDM model is the existence of plenty of metal-rich stars ([Fe/H] $\approx$ 0) and/or disk-like orbits with an estimated age of $\gtrsim$ 13 Gyr. Thus, these recent studies on the topic of the age of the Universe may allow higher limits, consistent with the opt-$\nu$HDM model ($t_0 \approx 15$ Gyr). 
%One thing to bear in mind, though, is that $H_0$ data or prior should not be input into the analysis whose aim is examine the value of $H_0$ \citep{2023arXiv230109695L}. % SG it is not clear what the point of this statement is! How made the mistake linder talks about?

The $\omega_{\nu}$ parameter is designed to be $\gg \omega_{\text{CDM}}$ in the neutrino-dominated models. In the opt-$\nu$HDM model, the sterile neutrino density $\omega_{\nu}$ replaces the cold dark matter density $\omega_\text{CDM}$ in driving the varying peak heights of the CMB power spectrum, an assumption that \citet{Witt} made. Nonetheless, we find that the best fitting $\omega_{\nu}$ is increased by $\approx 8.5 \%$ compared to the $\nu$HDM model. This means that the sterile neutrino mass will increase to $\approx$ 13 eV/c\textsuperscript{2}. This is significant though in the sense of the free-streaming length of the sterile neutrino, which is mass-dependent. In the opt-$\nu$HDM simulations, the sterile neutrino is therefore expected to cluster on larger scales than in the $\nu$HDM model. A possible way to examine this is the resulting neutrino fraction of the halos of baryons and neutrinos. The baryonic part stays constant, but the total matter contribution ends up being $\Omega_m \approx 0.49$. Because of the small Hubble parameter value, this results in a very large matter density. Such heavy Universe will have a rather different background cosmology, an extended matter-dominated era, and a shorter $\Lambda$ epoch domination.  

The spectral index $n_s$ describes how the density fluctuations vary with scale. If $n_s = 1 \approx n_{s}^{{\Lambda}\text{CDM}} \approx 0.966$, then the variations are the same on all scales ($\forall k = 2\pi$/$\lambda \rightarrow k^{\text{comoving}} = 2\pi \alpha \text{/}\lambda$, where $\alpha$ is the scale factor). Generically, cosmological inflation predicts $P \left( k \right) \propto k^{n_s-1}$, so that the matter power spectrum $P \left( k \right)$ is (almost) scale invariant. Our result deviates significantly from the Harrison-Zeldovich-Peebles approximation \citep{1970PhRvD...1.2726H, 1972MNRAS.160P...1Z, 1970ApJ...162..815P} of $n_s \approx $ 1 $\neq n_{s}^{\text{opt-}\nu\text{HDM}} \approx $ 0.872. A slight variation of the spectral index, $n_s$, would leave an imprint on the CMB anisotropies, since it is affecting the exponential potential of the accelerated expansion of inflation. For example, if $n_s$ is 10\% lower ($n_s \approx 0.78$), CAMB outputs an opt-$\nu$HDM CMB power spectrum which is shifted vertically, no longer fitting the \textit{Planck} data. The first peak will end up at $\ell \left( \ell \text{+}1 \right) C^{TT}_{\ell}\text{/}2\pi\approx$ 6342 [$\mu$K\textsuperscript{2}], while the rest of them will be moved to lower $\ell \left( \ell \text{+}1 \right) C^{TT}_{\ell}\text{/}2\pi$ values (y-axis of Fig. \ref{cmb}). The corresponding peaks will not be shifted either leftwards or rightwards; all of them will stay at the same $\ell$. From eq. \ref{pkformula}, we can see that the power spectrum behaves as a power law, but now it varies with scale, inducing a first-order variance, apart from the second-order corrections of $d n_s \text{/} d \ln k \sim \BigOSI[\big]{0.0001}{}$. Note also that the scale-variance does not refer yet to tensor perturbations, which are lower-order effects and very low values of $\ell$ in the CMB spectrum ($2<\ell<50$, Fig. \ref{cmb}) \citep{1995PhRvD..52.1739K}. In any case, this departure from scale invariance, will effectively categorize the opt-$\nu$HDM as a \textit{tilted} model, forcing a characteristic scale of the power spectrum, where some perturbations grow more and others not at all. This naturally leads to an inequality of power and will eventually be seen during the structure formation. Speculatively, this is a bad sign for the opt-$\nu$HDM model since there is already a contradiction from the galaxy cluster mass function, seen by \cite{Witt}. However, before concluding, we should keep in mind that MOND's non-linear gravity will enhance these anisotropies and perhaps create the "desired" mass function of galaxies and galaxy clusters. Clearly, it is necessary to test this problem with numerical simulations.
%The power spectrum of the temperature fluctuation is defined as $C_{\ell} \equiv \frac{1}{2\ell\text{+}1} \Sigma^{\ell}_{m=-\ell}\langle |a_{lm}|^2 \rangle$, where $\ell$ is the multipole and represents a given angular scale in the sky $\alpha$, given approximately by $\alpha = \pi$/$\ell$ (in degrees) and $a_{lm}$ are the coefficients of the spherical harmonics expansion of the temperature mean deviation, following the standard \textit{Planck} notation.

The optical depth to re-ionization, $\tau$, is a unit-less quantity which provides a measure of the line-of-sight free-electron opacity to CMB radiation. It is a quantification of the era when emitting sources first began forming and re-ionizing the neutral gas that existed after recombination. A lower (than the standard $\Lambda$CDM) value would mean a lower number density of ionizing sources and longer "dark ages" for the opt-$\nu$HDM model. Moreover, a lower $z_{\text{reion}}$ and consequently a later onset of the re-ionization period (see Table \ref{posterior_mean}) would mean that most of the re-ionization of the Universe happened rather recently, $z<6$. Recall that for the $\Lambda$CDM model estimate, $z_{\text{reion}}\approx$ 8-6.

Last, the structure amplitude parameter $A_\text{s}$ is set by hand in CAMB for the $\nu$HDM model equal to $ A_\text{s}^{{\nu}\text{HDM}} = $ 2.100549e-09 = $A_\text{s}^{{\Lambda}\text{CDM}}$ $\neq A_\text{s}^{\text{opt-}\nu\text{HDM}} =1.982e-09$. It comes accordingly with the decrease of the spectral index and both parameters will produce rather non-trivial results for the primordial perturbations.

%Lastly, the $N_\text{eff}$ parameter marks the number of relativistic species in the young Universe. Typically, only active neutrinos contribute summing up to 3.044. The small shift of .044 is due to the slight deviation from a thermal distribution. \citet{Witt}, using CAMB, have modified the $N_{\text{eff}}$ to 4.048 due to the addition of the sterile neutrino and so we respected that too in the CosmoSIS namelist.

\subsection{Derived initial conditions for simulations}
Using the best-fit values in our different models, we can calculate the matter power spectrum and the transfer function, which are presented in Fig. \ref{pk} and Fig. \ref{transfer} respectively. We have reproduced the \citet{Witt} results and added the opt-$\nu$HDM ones, on top.  The green opt-$\nu$HDM P(k) is well-differentiated at the very small scales (high k), but behaves similarly at the large scales (low k), as the $\nu$HDM model. This is expected since neutrinos do not have any power at all at galaxy scales (P(k) $< 10^{-5}$ for $k>10^{-1}$ h/Mpc) and consequently do not clump on these. A slight decrease in the large-scale power is expected by the slightly higher $m_{\nu_{s}}$. However, the P(k) itself is not very illuminating, because it does not explicitly demonstrate the small-scale differences. Counting on the fact that Fig. \ref{pk} is a logarithmic scale, the slight differences on the free-streaming length of the neutrino between the $\nu$HDM and the opt-$\nu$HDM models are difficult to discriminate. Another point would be that the free-streaming length will also depend on the expansion of the model, since we will be considering the comoving length. And as we have seen, the opt-$\nu$HDM model is rather different from its progenitor $\nu$HDM model in the sense of the expansion history, due to the low $H_0$ value.

The transfer function describes the evolution of perturbations through the epoch of crossing the radiation to the matter-dominated era. The main described effects are the change of the growth of fluctuations from the radiation-dominated era to the subsequent matter-domination one, which occurs around $z\approx 3000$ in $\Lambda$CDM. In the opt-$\nu$HDM, the redshift will be approximately the same, but the energy densities $\Omega_i$ inside the Friedmann equation will affect the Hubble parameter, since the matter content will be greater. Another effect will come from the radius of the horizon $r_H\left( z \right)$, since matter can only interact on scales $k<r_H\left(z\right)$. Regarding the very large scales $k<k_0$, the transfer function stays flat till $k_0$, where $k_0$ is the turnover from linear to non-linear collapse which is the change of the behavior in Fig. \ref{pk} ($10^{-2}<k_0<10^{-1}$). For smaller scales $k>k_0$, perturbations smaller than the size of the horizon will evolve in the non-linear regime as $T \left( k \right) \propto \left( \frac{k_0}{k}\right)^2$ and consequently the Primordial power spectrum $P\left( k, t_0 \right) = A_\text{s} \times k^{n_s} \times T^2\left( k \right) \propto k^{-3}$. It is also important to be aware that there is a transfer function for each of the species, i.e. photons, baryons, cold dark matter and neutrinos, as well as transfer functions for their velocities, too. Here, we study the transfer function of the most dominant element of each of the models. That is, the transfer function of the CDM component of the $\Lambda$CDM model, the sterile neutrino of the $\nu$HDM model. We specifically plot the absolute value $|T\left(k\right)|$, since many modes contribute negatively and might be wrongly neglected. During a hydro simulation, some over-densities will turn into an under-density and vice versa. To capture this procedure, we plot the absolute value. In a different case, the $T \left( k \right)$ of the neutrinos would be even smaller than its absolute value outcome. So, looking at Fig. \ref{transfer}, the excessive green (to red) line has a vital role. It might cause the formation of even more massive clusters in the simulations. It will actually enhance the structure formation, possibly solving the late-time emergence of the cosmic web in the MOND $\nu$HDM cosmogony. It is, in any case, compelling to perform fresh opt-$\nu$HDM numerical calculations. All the plots bend towards higher k (low $\lambda$ wavelength) due to the M\'esz\'aros effect, which is a consequence of the early radiation domination. The CDM component does not have any oscillations due to its collision-less configuration. The small-scale acoustic oscillations are prominent in the neutrino-characterized models, which quantify the Silk damping. Both of the neutrino models die out at high $k$, as designed, the regime where MOND will take over, providing the gravitational effects. Importantly, though, the transfer function of every species depends on the gravitational potential (and its Fourier transform). We can still trust them though, since they refer to high redshift $z\approx199$, where all the accelerations are in the Newtonian regime and "far" enough from the time of MONDian low accelerations ($z\approx50$). Numerically, the transfer functions can be used by other packages, like MUSIC \citep{2011MNRAS.415.2101H} to generate Initial Conditions for N-body and hydro solvers, like the Phantom of RAMSES code \citep{2015CaJPh..93..232L}.

%\ns{Regarding the very large scales $k<k_0$, the transfer function scales as $|T\left( k \right)| > 1$, where $k_0$ is the turnover from linear to non-linear collapse which is the change of the behavior in Fig. \ref{pk} happening $10^{-2}<k<10^{-1}$. This is generally not shown in transfer function plots and would be visible only at the left or Fig. \ref{transfer}. For smaller scales $k>k_0$, perturbations smaller than the size of the horizon will evolve in the non-linear regime as $T \left( k \right) \propto \left( \frac{k_0}{k}\right)^2$ and consequently the Primordial power spectrum $P\left( k, t_0 \right) = A_\text{s} \times k^{n_s} \times T^2\left( k \right) \propto k^{-3}$. \linebreak }

\subsection{Comparison to distance measures}
\begin{figure*} 
  \includegraphics[width=\textwidth]{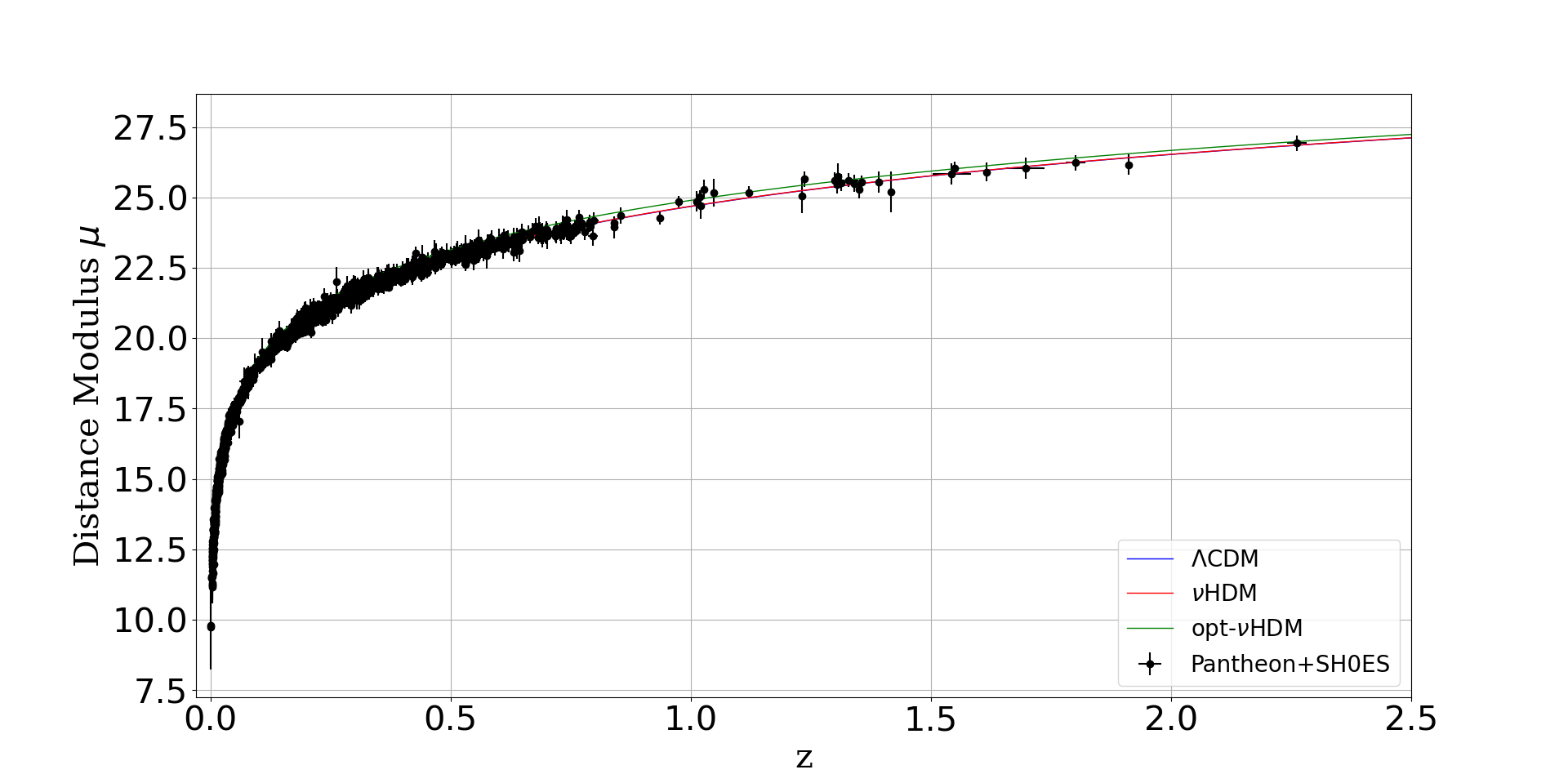}
  \caption{Distance Modulus showing data points taken from Pantheon+SH0ES.}
  \label{distance_modulus}
\end{figure*}

In this section, we compare our aforementioned models with observational data from Pantheon+ \citep{2022ApJ...938..110B} and SH0ES \citep{2022ApJ...934L...7R}. In Fig. \ref{distance_modulus}, on the distance modulus plot, one recognizes the closeness of the 3 models. One might further claim that the opt-$\nu$HDM model fits the data even better, especially at higher redshifts $z>1$, while at later times, the three theoretical scenarios are indistinguishable. The three studied cosmological models are within the range of uncertainties. 
%\ns{All the cosmological distances do not depend explicitly on the gravitational potential and are derived from an FLRW metric.}

Next, we study the cosmological distance $D_M$. CosmoSIS itself is able to calculate the distances. The most crucial parameter is the drag-epoch sound horizon $r_d$ and it is model-dependent. The $r_d$ physically quantifies the distance that sound waves can travel between the Big Bang and the drag epoch, which indicates the time when the baryons decoupled. Theoretically, in a $\Lambda$CDM Universe, the drag epoch occurs slightly later ($z_d =$  1060) than photon decoupling ($z_{\star} = $  1090) simply because there are so few baryons relative to the number of photons.

BAO measurements, which is what DESI \citep[section 2.1]{2025JCAP...02..021A} is probing, depend on the size of the sound horizon $r_d$, which mathematically is:
\begin{equation}
    r_{d} = \int_{z_d}^{\infty} \frac{c_s \left( z \right)}{H\left(z\right)}\,dz \text{ and } c_s\left( z \right) = \frac{c}{\sqrt{3 \big( 1 \text{+} \frac{3 \rho_{B}}{4\rho_{\gamma}} \big)}},
\end{equation}
where $H\left( z \right)$  is the Hubble parameter, $c_s$ is the speed of sound prior to recombination and $\rho_B$ and $\rho_{\gamma}$ are the baryon and radiation energy densities, respectively.

%Before inspecting the Fig. \ref{d_V}, we highlight the non-dependence of the physical quantity $D_V$ on the gravitational potential. This is not a hydro-dynamical simulation post-process study. In that case, one should pay attention to the arising of MONDian gravitational non-linearities. More analytically, $D_{V}$ is the angle-average distance that quantifies the average of the distances measured along, and perpendicular to, the line of sight to the observer:
%\begin{equation*}
 %   D_{V} \left( z \right) = \Big[ z \cdot D^2_M \left(z\right) \cdot D_H \left( z \right) \Big] ^ {\frac{1}{3}}, 
%\end{equation*}

Following DESI notation \citep{2025JCAP...02..021A}, in a FLRW metric, the transverse comoving distance is:
\begin{equation}
    D_M \left( z \right) = \frac{c}{H_0 \sqrt{\Omega_\text{K}}} \sinh \left[ { \sqrt{\Omega_\text{K}} \int_0^z \frac{dz^{\prime}}{H\left(z^{\prime} \right) \text{/}  H_0}]} \right] = \frac{r_d}{\Delta \theta},
    %\text{ and } D_H \left( z \right) = \frac{c}{H\left( z \right)}
\end{equation}
where $\Omega_\text{K}$ is the curvature parameter, defined as $\Omega_\text{K} = 1 - \Omega_m -\Omega_{\Lambda}-\Omega_\text{R}$ at the present day, $H_0$ is the Hubble constant value today, $\sinh$ is a the hyperbolic sinus function and last the Hubble parameter is :\linebreak $H\left( z \right) = \sqrt{ \Omega_{\Lambda} \text{+ } \Omega_\text{K} \left( 1 \text{+} z \right)^2 \text{+ } \Omega_m \left( 1 \text{+} z \right)^3 \text{+ } \Omega_\text{R}\left( 1 \text{+} z \right)^4 }$, \\
with z being the redshift. \linebreak
Non-relativistic massive neutrinos (sterile ones included) are counted as radiation before decoupling. DESI probes at low redshifts and massive neutrinos at that period are part of the matter budget though. The $\Omega$ parameters are the energy density values relative to the critical density of each of them at the present time. Using the DESI concept, for a group of pairs of galaxies at a given redshift, and a separation vector of the pairs perpendicular to the line-of-sight from the observer, the preferred angular separation is $\Delta \theta$, measuring the comoving distance $D_M\left( z \right) \equiv \frac{r_d}{\Delta \theta}$ at that redshift.

\begin{figure*} 
 \includegraphics[width=\textwidth]{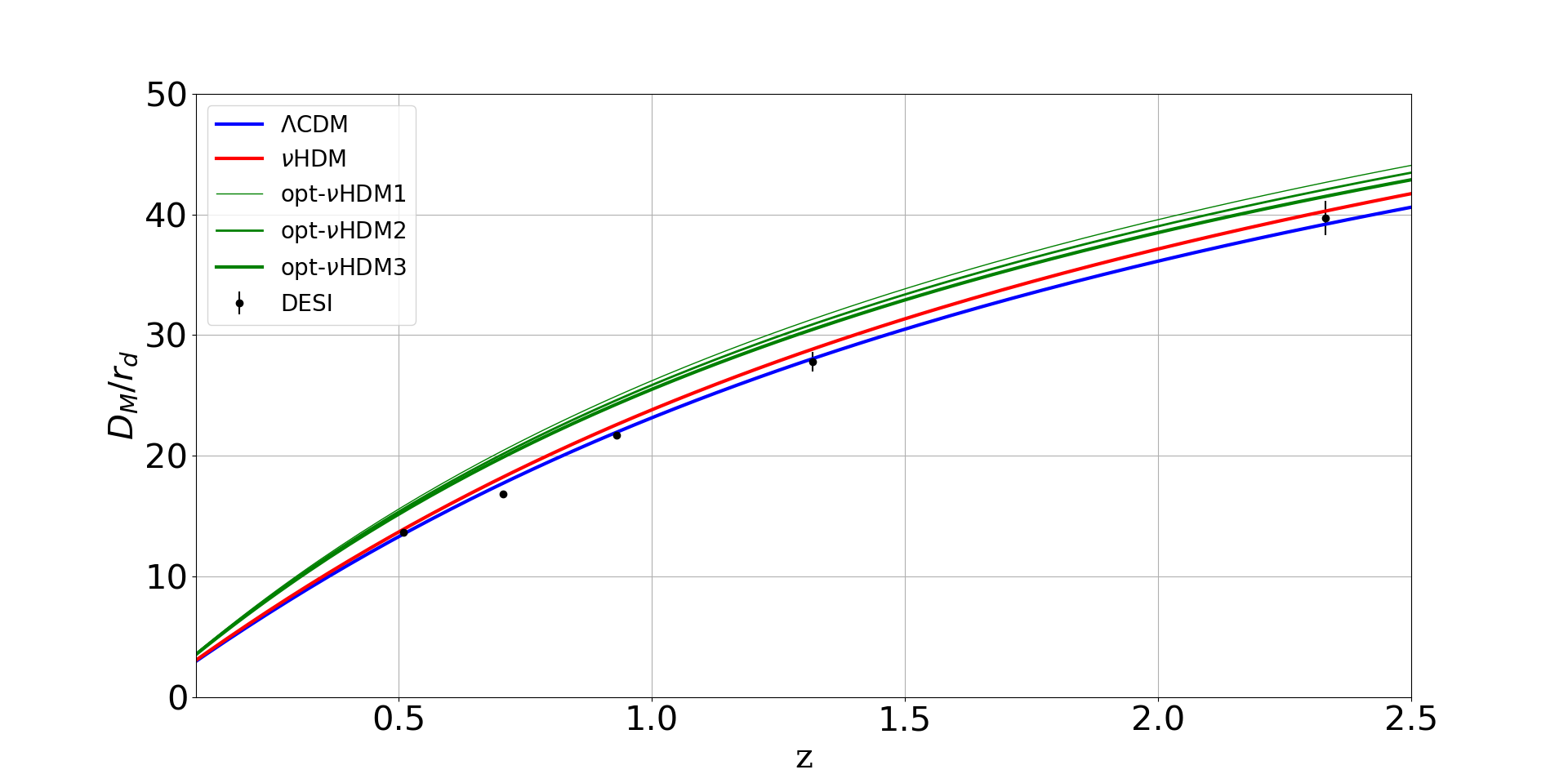}
  \caption{$D_{M}$. Three different green lines for the opt-$\nu$HDM, corresponding to opt-$\nu$HDM[1,2,3]->[130.05, 145.05, 160.05] Mpc.\\
  Data points are taken from DESI Collaboration \citep{2025JCAP...02..021A}}.
  \label{d_M}
\end{figure*}

%\begin{figure*} 
 % \includegraphics[width=\textwidth]{d_v.png}
 % \caption{$D_{V}$. Three different green lines for the opt-$\nu$HDM, corresponding to opt-$\nu$HDM[1,2,3]->[130.05, 145.05, 160.05] Mpc.\\
 % Data points taken form \citet{2024arXiv240403002D}}
 % \label{d_V}
%\end{figure*}

Nonetheless, CosmoSIS does not directly provide an $r_d$ value, contrary to CAMB. This is why, for the opt-$\nu$HDM model, we assume, as an ansatz, 3 different values of $r_d = \big[ 130.05,145.05,160.05 \big]$ Mpc. The standard $\Lambda$CDM value is 147.05 Mpc \citep[their equation 2.5]{2025JCAP...02..021A}. As for the opt-$\nu$HDM model, the sound horizon will probably be $r^{opt-\nu{\text{HDM}}}_d<145$ Mpc, because of the lower Hubble expansion and the higher $\Omega_m$, but this needs to be investigated. Despite the large variations around the standard $\Lambda$CDM value of $r_d=147$ Mpc, the opt-$\nu$HDM is not able to fit the $D_M$/$r_d$ DESI data, either with much lower ($r_d = 130$ Mpc) or with a much higher value ($r_d = 160$ Mpc). Note that the $D_M$/$r_d$ error bars are of the order of $\approx$1.5 Mpc (their section 4.1). Prematurely, the opt-$\nu$HDM may give inaccurate $D_M$/$r_d$ distance estimates, but the DESI calculations are strongly $\Lambda$CDM dependent and they explore $\Lambda$CDM variants, like $w_0 w_{\alpha}$CDM, considering solely the Newtonian potential at lower redshifts. In their section 3.1, \citet{2025JCAP...02..021A} mention that the y-axis is scaled by an arbitrary factor of $z^{-\frac{2}{3}}$ for visualization purposes, but in our analysis the results do not have any scaling.

As seen in Fig. \ref{d_M}, the opt-$\nu$HDM model $D_M$ predictions are not matching the DESI data, taken from \citep[their table 1]{2025JCAP...02..021A}. The opt-$\nu$HDM cosmology, continuously from low to high redshift, overestimates the $D_M$ distances. Nevertheless, the data points themselves are model-dependent and thus are not an apple-to-apple comparison. For a proper comparison, one must re-calculate the $D_M \text{/} r_d$ values, since any BAO conjecture would require a theoretical model in the background, which in the DESI case, is the $\Lambda$CDM model.
%This is a highlighted point in the DESI considerations too \citep[section 2.1]{2024arXiv240403002D}, going not too far from $\Lambda$CDM model \citep{Thepsuriya_2015}.}

%Interestingly, the opt-$\nu$HDM1 model is the closest one to the highest-z data point at $\approx$ 2.4. One notices that removing the arbitrary scaling, the matching of previous models is no longer present. It is only at the very late times, $\approx$ 0.5, when the $\Lambda$CDM and the $\nu$HDM models are correctly predicting the angular distances.

%On the distance modulus plot \ref{distance_modulus}, all the theoretical models are indistinguishable, matching the observational data points. One could actually favor both $\nu$HDM models for z >1. 
%On the contrary, on the $D_v$ plot \ref{d_V}, for late times, the data don't "prefer" any theory, but on earlier times, none of the theoretical scenarios match the observation

\subsection{DM models}
In order to tackle the missing mass/gravity problem in clusters (and/or galaxies), the neutrino hypothesis has been proposed a long time ago \citep{1980PhRvL..45.1980B}. Massive neutrinos (like our non-zero-mass sterile one) are part of a bigger class of Hot/Warm DM models. What differs in our approach is the gravity, from Newtonian to MONDian. For example,  \citet{1996ApJ...458....1C}, after performing hydro simulations with massive neutrinos, concluded that a Warm DM approach produces too many rich clusters, just like \citet{Witt}, but working with standard Newtonian gravitation.  Moreover, based on standard Newtonian gravity, \citet[their section 4.2]{2017JCAP...11..046M} found that their non-cold DM models predict a lower abundance of halos/galaxies in the mass range of $10^8-10^9 M_{\odot}$, similarly to the \cite{Witt} findings. These numerical experiments indicate that maybe this N-body approach (Newton or Milgrom) is insensitive either to the gravity law or to the ICs. Alternatively, the enigma lies in the background evolution, which is governed by the metric and the cosmological parameters (energy budget on each of $\Omega_{\text{CDM}}, \Omega_{\nu}, \Omega_\text{K}$ and $\Omega_{\text{b}}$).

Concerning the neutrino production mechanism, numerous suggestions exist in the literature. However, some proposals originate from particle physics, while others come from cosmology, which can sometimes be incompatible.  As an example, \citet{PhysRevD.88.043502} had shown that a non-resonant transition resulting from a primordial lepton asymmetry is incompatible with the standard structure formation scenario. Non-standard BB nucleosynthesis (BBN) scenarios are suggested to investigate this, using software like PArthENoPE \citep{2022CoPhC.27108205G}. Note also that different production mechanisms have been proposed, relying for example, on a hypothetical particle in the early Universe that can result in a sterile neutrino. See, for instance, a singlet scalar particle interacting with the Higgs boson \citep{König_2016}. But again, a note of caution must be taken when BBN results are compared to CMB experiments, since most of the measurements are model-dependent and do not consist of an accurate comparison.

%See the difference between the squared terms $V^{L}$ and $V^{T}$ on the denominator of equation 6.6 from \citep{2017JCAP...01..025A}. commented out, since referree didnt like 11 Dec2024

\subsection{Further improvements} \label{corrections}
Last, attention must be given to the incorporated foreground and background physics \textit{Planck} type experiments are treating. The published power spectrum is obtained from the observed CMB after subtraction of the foreground and after correcting for five physical processes that lead to flux variations on the present-day sky as the CMB photons propagate from the surface of the last scattering at $z \approx 1100$ to redshift $z=0$. The five processes are:
\begin{enumerate}
    \item The gravitational redshift of the photons due to differences in the matter density at the surface of last scattering (the non-integrated Sachs-Wolfe effect).
    \item The corrections to the photon energy due to the photon falling into potential wells that evolve as they leave the well (early-time and late-time gravitational redshifts, the integrated Sachs-Wolfe effect).
    \item The changes in the photon energy due to the photons scattering on electrons in the evolving plasma clouds as structure formation proceeds (the Sunyaev-Zeldovich effect).
    \item The changes of the photon propagation direction as structure grows, through weak-lensing.
    \item Hitherto not accommodated foreground sources of photons may exist from very early galaxy formation \citep{Haslbauer_2022, 2025NuPhB101716931G}.
\end{enumerate}

These corrections are done based on the observed dusty map and the multi-frequency decomposition of the \textit{Planck} data, which assume a $\Lambda$CDM Universe. The structure growth in the opt-$\nu$HDM model is governed by Milgromian gravitation, whose non-linear nature, will change extensively the formation and evolution of large-scale structure down to galaxies. The exact scale of non-linear collapse in a MONDian simulation, the time-variant gravitational potential and the expansion of space will be different in MOND and hence all the decomposition sky maps need to be re-analyzed. Recent weak-lensing analysis \citep{2024ApJ...969L...3M} report that circular velocities of isolated galaxies are flat out to 1 Mpc, being consistent with a MONDian logarithmic gravitational potential, with no sign of DM, which is the basis of $\Lambda$CDM model.

Hydro-dynamical simulations are thus necessary to enlighten how perturbations grow in the opt-$\nu$HDM model, but the \textit{Planck}-derived CMB power spectrum might not be a valid constraint for non-$\Lambda$CDM models of structure formation. Based on the resulting parameters of the simulations, one should work backwards to re-calculate the power spectrum from the observed all-sky flux, by applying the corrections enumerated above.  In order to estimate self-consistently the CMB foregrounds, one must follow an iterative process of re-modeling and re-fitting the temperature fluctuation power spectrum, determining the difference to the $\Lambda$CDM model. Unbiased corrections according to MOND are decisive in order to derive robust conclusions for the $\nu$HDM models.

\section{Conclusions} \label{conclusions}
We have extended the study of the only-known cosmological model based on MOND, for which, structure formation simulations exist, the so-called $\nu$HDM model. The latest structure formation simulations were performed by \citet{Witt}. Milgromian Dynamics has been outstanding, predicting how galaxies function, from the flattening of rotation curves at the outskirts of the galactic systems to the tight relation between the observed and the theoretically estimated gravitational acceleration (RAR). However, MOND under-predicts the dynamical mass of galaxy clusters \citep{1999ApJ...512L..23S} and is not able to provide ICs which are required to test the validity of the gravitational force via numerical simulations.
%\ns{and we have tried to make a more "modern" constrain of the cosmological parameters governing its evolution, based on \textit{Planck} satellite information.}

Taking into account the difficulty MOND faces at the scale of groups and clusters of galaxies and the need for realistic ICs, we optimally fitted the CMB power spectrum for the $\nu$HDM cosmological model, given the latest \textit{Planck} data, re-naming it to opt-$\nu$HDM model. Using Bayesian statistics, we estimated the best-fitted values of the cosmological parameters which are strikingly different than the last observational studies of \citet{2011ApJ...730..119R} and DES \citep{PhysRevD.105.043512}. The central outcome of the Bayesian analysis is the low $H_0 \approx$ 55.65 km/s/Mpc and the heavy $\Omega_m \approx 0.5$. A premature conclusion on this cosmological scenario would be to exclude the $\nu$HDM model. The attempts of \citet{1984ApJ...286....3F} and \cite{1998MNRAS.296.1009S} to produce a solid MOND cosmology, which led to this concept of the complementary sterile neutrino, might come to an end. Nevertheless, one should consider novel observational investigations for stars with ages older than the age of the $\Lambda$CDM Universe \citep{2024arXiv240718307N, 2023eppg.confE.231K}. 

Therefore, in order to thoroughly investigate the resulting parameters, one should conduct hydro-dynamical numerical simulations to study the early-galaxy formation at $z\approx 10$ and the "correct" galaxy-cluster mass function. These efforts are already under-way (Samaras et al. 2025 - in preparation), but one thing to deal with is the higher $\Omega_{m} \approx$ 0.5, than the $\Omega_{m}^{\text{$\Lambda$CDM}} \approx$ 0.3 value, which will make the integrations much heavier and possibly the need of supercomputers.   

Last thing to bear in mind is instinctively the initial conditions of all the aforementioned models, which are in effect similar; all the matter (CDM/HDM+baryons+neutrinos) is created in a primordial singularity (BB). On top of that, inflation is postulated to make causally disconnected regions of the CMB homogeneous in a flat Euclidean Universe. The CMB is essentially used as a boundary condition for calculations. Moreover, the Cosmological Principle is encapsulated in the FLRW metric, which is the backbone of all CDM/HDM models. Although the CMB is very precisely measured and corrected, it might not be of significance in a MONDian cosmology. Caution must be taken when other theoretical models are compared to $\Lambda$CDM-calibrated data. CMB foregrounds and distance measurements are processed through the standard $\Lambda$CDM model, hence a legitimate comparison is not existent.
%By construction, these theoretical models vary only quantitatively, but not qualitatively, both resulting with minor and major inconsistencies with observational data of different scales. 

% Instead of by hand trying to match the CMB with a large number of parameters (like on the CAMB namelist)

%\section*{Appendix} \label{appendix}
%\begin{figure*} 
 % \includegraphics[width=0.9\textwidth, height=11cm]{corner.png}
 % \caption{Corner plot of the posterior sample}
 % \label{corner}
%\end{figure*}

%\begin{figure*} 
 % \includegraphics[width=\textwidth, height=8cm]{pk_2d.png}
  %\caption{Matter Power Spectra}
  %\label{pk_2d}
%\end{figure*}

\begin{table}
\centering
    \begin{tabular}{llll}
        $\left( \ell, \frac{\ell \left( \ell \text{+}1 \right) C^{TT}_{\ell}}{2\pi} \right)$ & \textit{Planck}   & $\nu$HDM & opt-$\nu$HDM \\ \hline
        1st Peak  & (225,  5793)   &  (222,  5944)  &  (221, 5752)  \\
        2nd Peak   & (524,  2573)   &  (539,  2825)  &  (537, 2580)  \\
        3rd Peak   & (824,  2522)   &  (822,  2617)  &  (814, 2524)  \\
        4th Peak   &(1124,  1232)   & (1136,  1396)  & (1128, 1321) \\
        5th Peak   &(1424,   809)   & (1441,   830)  & (1429,  800) \\
        6th Peak   &(1724,   397)   & (1751,   434)  & (1738,  420)  \\
    \end{tabular}
    \caption{$\left( \ell, \frac{\ell \left( \ell \text{+}1 \right) C^{TT}_{\ell}}{2\pi} \right)$ values for different models. \linebreak Note that the opt-$\nu$HDM values are computed with CosmoSIS, while for the $\nu$HDM they are calculated with CAMB. The \textit{Planck} values are from \citet{2020A&A...641A...1P}.}
    \label{peaks}
\end{table}

%%%%%%%%%%%%%%%%%%%%%%%%%%%%%%%%%%%%%%%%%%%%%%%%%%%%%%%%%%%%%%%%%%%%%%%%%%%%%%%%%%%%%%%%%%%%%%%%%%%%%%%%%%%%%%%%%%%%%%%%%%%%%%%%%%
\section*{Acknowledgments}
NS is supported by the AKTION grant number MPC-2023-07755. Most of this work took place in Innsbruck, where he stayed for the 3 awarded months (March-May 2024). NS would also like to thank Prof. Francine Marleau and Prof. Tim Schrabback for their hospitality at the Institute for Astro- and Particle Physics of the University of Innsbruck. NS is also supported by the Charles University Grant Agency (GAUK) - 94224. NS and PK acknowledge support through the Deutscher Akademischer Austauschdienst (DAAD) Bonn-Prague exchange program. The authors would like to thank the anonymous referee for enhancing the quality of the manuscript.
%%%%%%%%%%%%%%%%%%%%%%%%%%%%%%%%%%%%%%%%%%%%%%%%%%%%%%%%%%%%%%%%%%%%%%%%%%%%%%%%%%%%%%%%%%%%%%%%%%%%%%%%%%%%%%%%%%%%%%%%%%%

%%%%%%%%%%%%%%%%%%%%%%%%%%%%%%%%%%%%%%%%%%%%%%%%%%%%%%%%%%%%%%%%%%%%%%%%%%%%%%%%%%%%%%%%%%%%%%%%%%%%%%%%%%%%%%%%%%%%%%%%%%%%%%%%%%
\section*{Data Availability} \label{data}
We acknowledge the use of the \textit{Planck} Legacy Archive. \textit{Planck} (\href{https://www.esa.int/Science_Exploration/Space_Science/Planck}{http://www.esa.int/Planck}) is a European Space Agency (ESA) science mission with instruments and contributions directly funded by ESA Member States, NASA, and Canada.

We make use of the publicly available data from Pantheon+ and SH0ES from their continuously updated web page, \href{https://pantheonplussh0es.github.io/}{Pantheon+SH0ES}.
We also made use of the \href{https://www.desi.lbl.gov/}{DESI Survey}, which is being conducted on the Mayall 4-meter telescope at Kitt Peak National Observatory. DESI is supported by the Department of Energy Office of Science to perform this Stage IV dark energy measurement using baryon acoustic oscillations and other techniques that rely on spectroscopic measurements. 
%The data points on the fig. \ref{d_M} are taken from the DESI pre-print table 1, which was published on the ADS on April 2024. 
\par The CosmoSIS used namelist files and the subsequent data are available under reasonable requests.
%%%%%%%%%%%%%%%%%%%% REFERENCES %%%%%%%%%%%%%%%%%%%%%%%%%%%%%%%%%%%%%%%%%%%%%%%%%%%%%%%%%%%%%%%%%%%%%%%%%%%%%%%%%
\bibliographystyle{mnras}
%\bibliography{BIBLIO}
%\bibliographystyle{plain}
\bibliography{BIBLIO}
\bsp	% typesetting comment
\label{lastpage}
\end{document}